\documentclass[manuscript]{acmart}
\usepackage{xcolor}
\usepackage{soul}
\usepackage[table]{xcolor}

\usepackage{tabularx}
\usepackage{stfloats}
\usepackage{booktabs}
\usepackage{array}
\usepackage{graphicx}
\usepackage{pifont}
\usepackage{float}
\usepackage{placeins}
\usepackage{multirow}
\usepackage[most]{tcolorbox}
\usepackage{pdflscape}\usepackage{adjustbox}
\usepackage{makecell}
\usepackage{pifont}
\usepackage{ragged2e}
\usepackage{tabularx}
\usepackage{longtable}
\usepackage{booktabs}
\usepackage{array}
\newcolumntype{L}[1]{>{\raggedright\arraybackslash}p{#1}}
\newcolumntype{C}[1]{>{\centering\arraybackslash}p{#1}}

\newcommand{\iconsize}{0.75ex}
\newtcolorbox{examplebox1}{
  colback=yellow!5,
  colframe=black,
  boxrule=0.4pt,
  arc=1pt,
  boxsep=0.5pt,
  left=2pt,
  right=2pt,
  top=1pt,
  bottom=1pt
}
\newtcolorbox{examplebox2}{
  colback=red!5,
  colframe=black,
  boxrule=0.4pt,
  arc=1pt,
  boxsep=0.5pt,
  left=2pt,
  right=2pt,
  top=1pt,
  bottom=1pt
}
\usepackage{tikz}

\AtBeginDocument{%
  }

\acmJournal{CSUR}


\begin{document}\hypersetup{
  colorlinks=true,
  citecolor=blue,
  linkcolor=blue,
  urlcolor=blue
}
\title{Shifting from Injection to Interaction: Rethinking Web Security in the Age of LLMs and Beyond}

\author{Nivedita Singh}
\email{singhnivvy@g.skku.edu}
\orcid{0000-0001-6225-3669}
\affiliation{
  \department{Department of Computer Science}
  \institution{Sungkyunkwan University}
  \streetaddress{Seobu-ro, Jangan-gu}
  \city{Suwon}
  \country{South Korea}
  \postcode{2066}
}

\author{Alsharif Abuadbba}
\affiliation{
  \institution{CSIRO}
  \country{Australia}
}

\author{Yansong Gao}
\affiliation{
  \institution{The University of Western Australia}
  \country{Australia}
}

\author{Surya Nepal}
\affiliation{
  \institution{CSIRO}
  \country{Australia}
}

\author{Hyoungshick Kim}
\email{hyoung@skku.edu}
\authornotemark[1]
\affiliation{
  \department{Department of Software}
  \institution{Sungkyunkwan University}
  \streetaddress{Seobu-ro, Jangan-gu}
  \city{Suwon}
  \country{South Korea}
}

\renewcommand{\shortauthors}{Singh, et al.}
\begin{abstract}
Large language models (LLMs) are becoming integral to web applications and browser agents, transforming online interactions while introducing new attack vectors and reshaping longstanding web vulnerabilities.
Classical threats such as cross-site scripting (XSS) can be amplified through LLM-mediated interactions, while LLM-specific vulnerabilities can propagate across web applications, introducing attacks such as prompt injection.
Securing modern web systems therefore requires understanding interactions between traditional and LLM-specific threats across the system lifecycle.
Unlike prior surveys treating web and LLM security separately, this survey provides a unified analysis of how LLMs amplify web vulnerabilities across client-side, server-side, and pipeline layers while evaluating defenses and their limitations.
The analysis examines extending NIST and ISO/IEC AI security frameworks to the security needs of LLM-enabled web environments.
Three unresolved challenges are identified: adversarial natural-language instructions, autonomous agent security, and post-deployment security through continuous monitoring and adaptation.
An LLM-aware monitoring and control framework is proposed, integrating semantic input validation, prompt integrity protection, output isolation, agent governance, and runtime monitoring.
This unified perspective characterizes the evolving threat landscape and outlines future directions for secure AI-enabled web systems.

\end{abstract}
\maketitle
\section{Introduction}
Whether automating software development tasks or generating multimedia content, large language models (LLMs) have become indispensable tools that can process natural-language prompts and deliver rapid, highly capable results~\cite{xu2024can}. LLMs are fundamentally reshaping how users interact with online services~\cite{sun2025friendly} through their rapid integration into web applications and browser agents. While their stunning capabilities offer significant benefits in terms of automation, personalization, and accessibility, they also introduce new challenges by drastically altering the web security landscape~\cite{pankajakshan2024mapping, kong2025survey, gan2026navigating, wu2024new}. 

\begin{figure}[t]
    \centering
     \includegraphics[width=0.5\columnwidth]{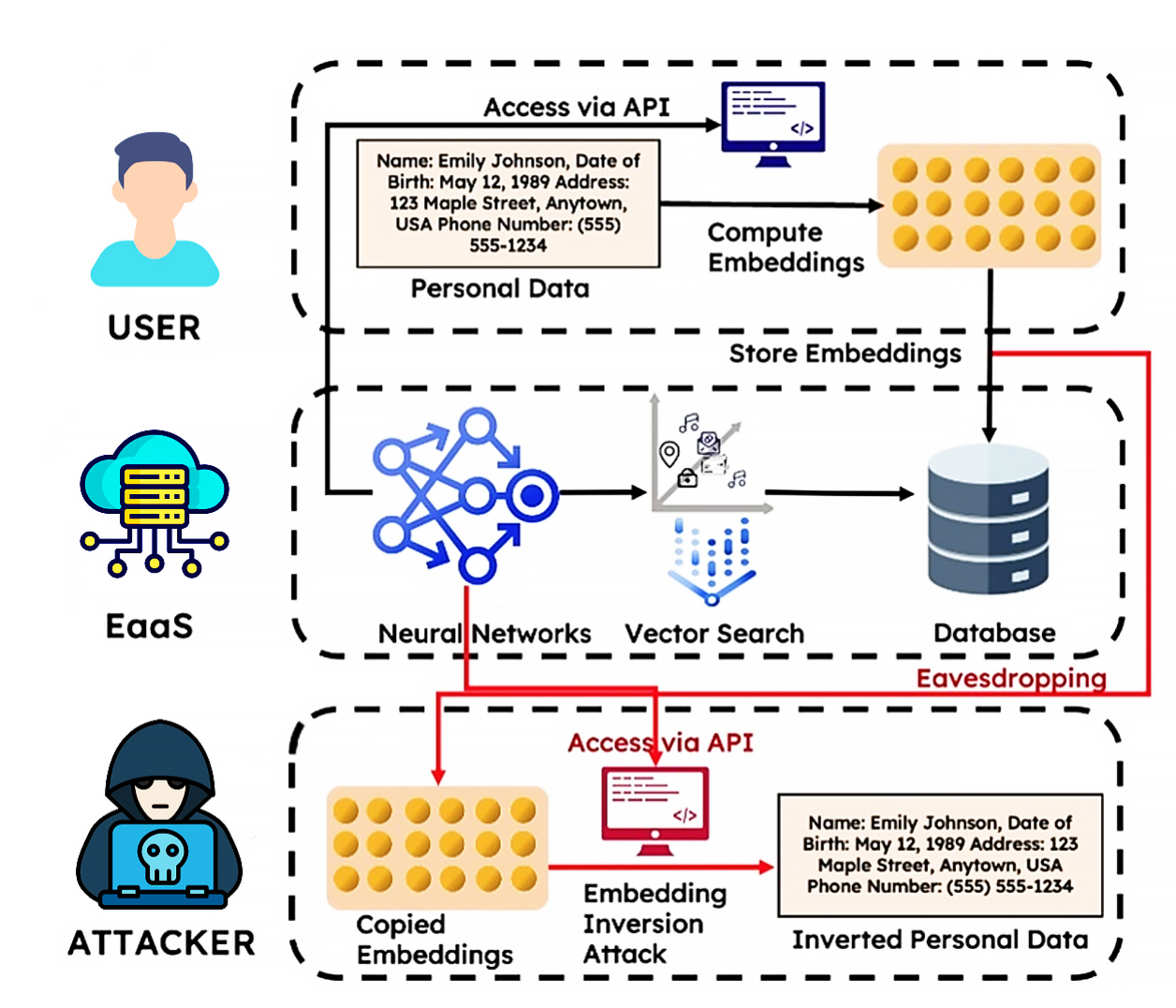} 
    \caption{Architecture of an embedding inversion attack on a language model API. By eavesdropping on vector search queries, an adversary can reconstruct sensitive personal data from intercepted embeddings, illustrating how backend pipeline vulnerabilities directly compromise client-level privacy. Adapted from Chen et al.~\cite{chen2024text} under CC BY 4.0; icons and presentation modified by the authors.}
    \label{fig:text_embeddings}
\end{figure}

Among cybersecurity domains, integrating client-facing interfaces with backend LLM pipelines has emerged as a critical security concern. As LLMs interact directly with end-users, their advanced capabilities transcend traditional security boundaries, allowing long-standing attack vectors such as XSS~\cite{gupta2019evaluation, rodriguez2024personal} and CSRF~\cite{peguero2021csrf} to evolve beyond their conventional forms. Rather than being isolated exploits, these legacy threats are now amplified and transformed through LLM-mediated interactions, significantly expanding the attack surface for adversaries. This emerging shift introduces entirely new attack paradigms. For example, prompt injection attacks can mimic the effects of XSS~\cite{mchugh2025prompt}, enabling unauthorized access to sensitive user data, while LLM-driven agents may unknowingly execute adversarial instructions embedded in web content. Beyond these adaptations, novel threats such as membership inference, jailbreaking, and model inversion further expose sensitive data in ways that web security frameworks are ill-equipped to handle. For instance, vulnerabilities in backend embedding pipelines can expose user data to inversion attacks (see Figure~\ref{fig:text_embeddings}), bridging server-side weaknesses with severe user-level privacy breaches.
Although the user interacts only with a visible web application, the security failure stems from interactions among multiple layers: the client-facing interface, the embedding model, the vector database, the retrieval API, and the downstream application logic. This scenario is grounded in the well-known problem of excessive data exposure (EDE), where web API transmits more information than the client interface legitimately needs~\cite{pan2024edefuzz}. This emphasizes that LLM-related vulnerabilities should not be analyzed solely at the model level, as backend pipeline weaknesses can propagate into concrete harms at the client side and in web applications.
Despite growing awareness of these risks, there remains a lack of systematic analysis that connects LLM-specific threats to concrete full-stack web attack models and mitigation strategies. To contextualize these risks, the Open Web Application Security Project (OWASP)  published a dedicated Top 10 list for LLMs and generative AI applications of 2025~\cite{OWASP_LLM}. As the \textit{de facto standard} for web application security, this community-driven effort extends OWASP’s long-standing tradition of identifying critical risks, focusing specifically on vulnerabilities introduced throughout LLM development, deployment, and integration. OWASP provides a structured lens for analyzing how LLMs reshape both legacy and novel security challenges. Building on this foundation, this work addresses two key research questions:

\textbf{RQ1.} How and to what extent do LLM security risks amplify, transform, or re-enable traditional web vulnerabilities across the web application stack?

\textbf{RQ2.} Are existing defenses and security standards sufficient to mitigate these LLM-mediated risks across client-side, server-side, pipeline, and ecosystem layers?

Existing works primarily analyze LLM vulnerabilities at the model or agent level, often treating LLMs as the final attack surface. However, this perspective overlooks a critical shift: LLMs act as intermediaries that propagate and amplify legacy web vulnerabilities across client–server boundaries. This gap motivates our study.
In this paper, we examine how LLM security risk manifests in practice from a web application's perspective, identifying both overlaps with and deviations from traditional full-stack attack models (RQ1). We then investigate how LLM integration re-enables long-standing web vulnerabilities, including XSS, CSRF, clickjacking, man-in-the-middle attacks, client-side request hijacking, and session hijacking, while also introducing novel risks—such as system-prompt leakage, embedding inference, misinformation, and excessive agency—that fall outside conventional web-security frameworks (RQ2). As part of RQ2, we further analyze how established security and risk management frameworks should be extended to systematically address these interconnected LLM-specific threats.
By embedding these RQs into our analysis, this survey highlights the urgent need to rethink application-centric defenses in the era of LLM-enabled web systems. Specifically, we make four key contributions:
\begin{itemize}
    \item First, we present a structured taxonomy that maps web attacks to their emerging manifestations under the OWASP LLM Top 10 risks, clarifying how LLM-specific designs create new web attack surfaces. 
    \item Second, we analyze how LLM-driven automation and natural language interfaces amplify and transform legacy vulnerabilities into subtler and more scalable attack vectors. 
    \item Third, we explore the mitigation strategies spanning client-side, server-side, and LLM pipeline components, exposing critical defense gaps that arise when web security controls are applied to LLM-integrated environments. 
    \item Finally, we discuss how existing security frameworks must evolve to account for LLM-driven ecosystem threats and outline actionable research directions toward secure and privacy-preserving LLM-enabled web ecosystems.
\end{itemize}

The remainder of this paper is organized as follows. 
Section~2 compares our work with recent LLM security surveys and identifies the gap that motivates a full-stack web security perspective. 
Section~3 presents our methodology. 
Section~4 describes the systematic literature review process, covering the search strategy, keywords, inclusion/exclusion criteria, and primary-study selection. 
Section~5 presents the comparative security analysis, showing how OWASP LLM risks map to traditional web vulnerabilities, how these risks are amplified in LLM-integrated web applications, and what defenses and open challenges remain. 
Section~6 discusses extensions to the security framework for LLM-enabled systems, and Section~7 concludes the paper.
\section{Comparison with Prior Recent Surveys}
Recent studies (2024--2026) have increasingly focused on systematizing the security and privacy risks associated with LLMs. As highlighted in Table~\ref{tab:PriorSurveys}, while several comprehensive surveys exist, they predominantly analyze LLM vulnerabilities in isolation or focus strictly on standalone agent interactions, largely overlooking the cascading effects on the broader web ecosystem.

\newcommand{\Full}{\tikz\fill[black] (0,0) circle (\iconsize);}
\newcommand{\Half}{%
\tikz[baseline=-0.6ex]{
    \fill[white] (0,0) circle (\iconsize);
    \fill[black] (0,0) -- (90:\iconsize) arc (90:270:\iconsize) -- cycle;
    \draw[black] (0,0) circle (\iconsize);
}%
}
\newcommand{\Empty}{\tikz\draw[black] (0,0) circle (\iconsize);}

\newcommand{\cmark}{\ding{51}}
\newcommand{\pmark}{\textcolor{black}{\ding{51}}}
\newcommand{\xmark}{--}

\definecolor{headerblue}{RGB}{220,235,247}
\definecolor{lightgrayrow}{RGB}{248,248,248}
\definecolor{ourrow}{RGB}{232,245,233}

\begin{table*}[t]
\centering
\caption{Comparison of Prior LLM Security Surveys with Our Full-Stack Web-Security Perspective. Filled circles indicate thorough coverage, half-filled circles indicate partial or brief coverage, and empty circles indicate no discussion. Checkmarks indicate whether a survey explicitly addresses the corresponding full-stack web-security dimension.}
\label{tab:PriorSurveys}

\scriptsize
\setlength{\tabcolsep}{2.2pt}
\renewcommand{\arraystretch}{1.12}

\resizebox{\textwidth}{!}{%
\begin{tabular}{L{3.2cm}
                *{10}{C{0.35cm}}
                C{0.75cm} C{0.75cm} C{0.85cm} C{0.85cm} C{0.75cm}}
\toprule
\rowcolor{headerblue}
\textbf{Prior Work} &
\textbf{01} & \textbf{02} & \textbf{03} & \textbf{04} & \textbf{05} &
\textbf{06} & \textbf{07} & \textbf{08} & \textbf{09} & \textbf{10} &
\textbf{Web} & \textbf{TRA} & \textbf{Layer} & \textbf{FW} & \textbf{Stack} \\
\midrule

Gan et al.~\cite{gan2026navigating} &
\Half & \Full & \Empty & \Full & \Half & \Full & \Full & \Half & \Empty & \Half &
\pmark & \xmark & \pmark & \pmark & \xmark\\

Tamuka et al.~\cite{tamuka2026securing} &
\Full & \Full & \Half & \Half & \Full & \Full & \Full & \Empty & \Half & \Full &
\pmark & \xmark & \pmark & \pmark & \xmark \\

He et al.~\cite{he2025emerged} &
\Full & \Half & \Empty & \Half & \Half & \Empty & \Empty & \Full & \Full & \Empty &
\pmark & \xmark & \pmark & \xmark & \xmark \\

\rowcolor{lightgrayrow}
Nie et al.~\cite{nie2024text} &
\Empty & \Empty & \Empty & \Empty & \Empty & \Empty & \Empty & \Full & \Empty & \Empty &
\xmark & \xmark & \xmark & \xmark & \xmark \\

Wu et al.~\cite{wu2024new} &
\Full & \Full & \Half & \Half & \Full & \Half & \Half & \Half & \Half & \Half &
\pmark & \xmark & \pmark & \pmark & \xmark \\

Chhabra et al.~\cite{chhabra2025agentic} &
\Full & \Half & \Half & \Half & \Full & \Full & \Half & \Half & \Half & \Half &
\pmark & \xmark & \pmark & \pmark & \xmark \\

Liao et al.~\cite{liao2025attack} &
\Full & \Half & \Empty & \Full & \Half & \Empty & \Full & \Half & \Empty & \Half &
\xmark & \xmark & \xmark & \pmark & \xmark \\

\rowcolor{lightgrayrow}
Kong et al.~\cite{kong2025survey} &
\Full & \Full & \Half & \Full & \Full & \Full & \Half & \Half & \Half & \Half &
\pmark & \xmark & \pmark & \pmark & \xmark \\

Li et al.~\cite{li2025security} &
\Full & \Empty & \Half & \Half & \Full & \Empty & \Empty & \Empty & \Empty & \Half &
\xmark & \xmark & \xmark & \pmark & \xmark \\


Yan et al.~\cite{yan2024protecting} &
\Half & \Full & \Empty & \Half & \Empty & \Empty & \Full & \Empty & \Half & \Empty &
\xmark & \xmark & \xmark & \xmark & \xmark \\

\rowcolor{lightgrayrow}
Gallegos et al.~\cite{gallegos2024bias} &
\Empty & \Half & \Empty & \Empty & \Half & \Half & \Empty & \Empty & \Empty & \Empty &
\xmark & \xmark & \xmark & \xmark & \xmark \\

\midrule
\rowcolor{ourrow}
\textbf{Our work} &
\textbf{\Full} & \textbf{\Full} & \textbf{\Full} & \textbf{\Full} & \textbf{\Full} &
\textbf{\Full} & \textbf{\Full} & \textbf{\Full} & \textbf{\Full} & \textbf{\Full} &
\textbf{\cmark} & \textbf{\cmark} & \textbf{\cmark} & \textbf{\cmark} & \textbf{\cmark} \\

\bottomrule
\end{tabular}}
\vspace{2pt}

\begin{minipage}{0.98\textwidth}
\footnotesize
\textit{Note:} Columns 01--10 correspond to OWASP LLM risks LLM\_01--LLM\_10. 
Web = explicit treatment of LLM-integrated web applications; 
TRA (Traditional Threat-Reference Architecture) = mapping to traditional web threats or CWE; 
Layer = client/server/pipeline propagation; 
FW = security framework adaptation; 
Stack = unified full-stack security perspective. 
Filled marks indicate full coverage and half marks indicate partial treatment.
\end{minipage}
\end{table*}

\textbf{Agent-Centric and General Attack Surveys:} A significant portion of the existing literature focuses on the security of LLM-based agents and general attack-defense paradigms. For instance, He et al.~\cite{he2025emerged} and Gan et al.~\cite{gan2026navigating} explore the security, privacy, and ethical concerns of autonomous LLM agents, emphasizing risks like data poisoning, model inversion, and malicious misuse. 
Wu et al.~\cite{wu2024new} move beyond standalone LLM analysis by examining real-world LLM-based web systems composed of frontends, web tools, plugins, and sandbox components. Similarly, Chhabra et al.~\cite{chhabra2025agentic}
survey agentic AI systems that combine planning, tool use, memory, and autonomy, enabling them to interact with web, software, and physical environments with limited human supervision.
Similarly, Liao et al.~\cite{liao2025attack}, and Kong et al.~\cite{kong2025survey} categorize broad attack vectors---such as prompt injection and jailbreaking---and review corresponding mitigation strategies. While these surveys provide extensive overviews of LLM-native risks, they often treat the LLM as the final attack destination rather than a conduit. Consequently, they do not explore in depth how these vulnerabilities intersect with and amplify legacy web threats on the client side.

\textbf{Privacy and Component-Specific Analyses:} Other works narrow their focus to LLM specific privacy dimensions or pipeline components. Yan et al.~\cite{yan2024protecting} delve into active and passive data privacy leakage, while Nie et al.~\cite{nie2024text} specifically examine vulnerabilities inherent in text embedding processes. Gallegos et al.~\cite{gallegos2024bias} provide a detailed survey on bias and fairness in LLMs, focusing on evaluation metrics, datasets, and mitigation techniques. However, their analysis remains centered on LLM behavior and fairness concerns, rather than LLM-integrated web applications where backend model security risks can propagate into client-side and web-application harms.

\textbf{Traditional Web-Application Security Surveys}
Prior surveys on web applications security have extensively examined established vulnerabilities such as XSS, CSRF, SQL injection, clickjacking, session hijacking, and browser-side attacks \cite{li2014survey, deepa2016securing,prokhorenko2016web}. They provide mature taxonomies and mitigation strategies for traditional client–server applications but generally assume deterministic processing and clear trust boundaries. Thus, they do not capture how LLMs interpret untrusted natural-language content, mediate tools and APIs, or propagate backend weaknesses to client-side harms. Our work extends this foundation by examining how LLM integration transforms, amplifies, or re-enables these threats across the full application stack.

\textbf{Our Contributions in Context:} As demonstrated in Table~\ref{tab:PriorSurveys}, our work distinctively maps the web threat landscape against the comprehensive OWASP LLM Top 10. While prior surveys provide valuable insights, our research bridges three critical gaps in the literature. First, we shift the focus from isolated LLM vulnerabilities to the \textit{amplification of legacy web threats} when LLMs are integrated into web applications, detailing how LLM-mediated interactions exacerbate existing client-side security risks such as XSS and CSRF. Second, we synthesize mitigation strategies not just at the model layer but also comprehensively across the client-side, server-side, and ecosystem layers. Finally, unlike the existing literature, we emphasize the urgent need to extend and adapt established security and risk management frameworks (e.g., OWASP, ISO/IEC standards, and NIST AI RMF) to systematically govern risks in the emerging, interconnected LLM-integrated web ecosystem.

\section{Our Methodology} 
\begin{figure*}
    \centering
    \includegraphics[width=0.9\textwidth]{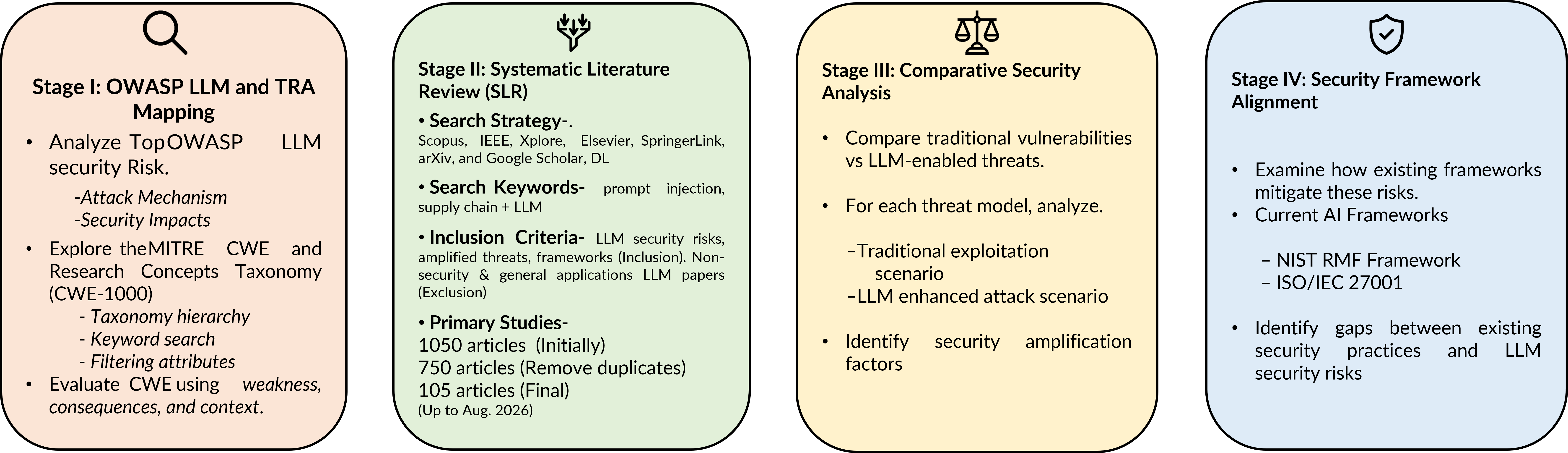} 
    \caption{Methodology pipeline for analyzing web-security vulnerabilities in LLM-enabled systems.}
    \label{fig:Metho}
\end{figure*}
Our study follows a structured multi-stage methodology (see Figure~\ref{fig:Metho}). We first analyze the OWASP LLM Top 10 risks to identify \textit{attack mechanisms and impacts}. To move beyond high-level risk categories toward root-cause weakness analysis, we further explore the MITRE CWE taxonomy (CWE-1000) using its \textit{hierarchy}, \textit{keyword search}, and \textit{filtering} capabilities to systematically identify relevant software and system weaknesses. We evaluate CWE entries based on their weakness descriptions, consequences, and operational context.
This initial OWASP--CWE analysis then informs the systematic literature review (SLR) by guiding the search keywords, inclusion criteria, and analysis of selected studies. Specifically, we use this mapping process to examine whether prior works discuss the corresponding LLM risk, its web-security analogue, affected system layer, and mitigation strategy. Following Kitchenham et al.~\cite{kitchenham2007guidelines} the SLR analyzes LLM security challenges, their amplification of traditional web vulnerabilities, and existing mitigations.


\subsection{\textbf{Web Security Frameworks and OWASP LLM Security Challenges}}

As the first step in our methodology, we analyze the OWASP LLM security risks and map them to corresponding TRA and CWE categories. This mapping requires a brief overview of the security frameworks and weakness taxonomies used in our analysis.

Security frameworks and standards~\cite{OWASP_LLM, OWASP_Traditional,CWE} identify and categorize security vulnerabilities in order to implement appropriate protection mechanisms and maintain secure web systems. These frameworks are periodically updated to address emerging and evolving security threats. 
Specifically, OWASP provides a community-driven, practical set of security challenges, such as the OWASP Top 10 LLM risks, that identify the most critical security risks in modern LLM-enabled applications~\cite{OWASP_LLM}. In contrast, CWE, maintained by MITRE, offers a standardized taxonomy of underlying software weaknesses with fine-grained technical definitions~\cite{CWE}. Together, OWASP and CWE enable a systematic view of security challenges, linking high-level threat categories to their root-cause vulnerabilities.
Additionally, there are also well-established security frameworks that provide mitigation strategies against such attacks, including standards developed by NIST and ISO~\cite{NIST, international2023iso}. 
These frameworks are built around the core principles of Identify, Protect, Detect, Respond, and Recover, which guide organizations in systematically managing and reducing security risks. 
Although these standards are globally recognized and widely adopted for managing cybersecurity risks, the emerging security threats introduced by LLM-enabled web systems are more dynamic and require stronger, LLM-specific mitigation strategies.

With the rapid emergence of generative AI, OWASP has published a dedicated Top 10 security challenges of the year 2025~\cite{OWASP_LLM} which we follow. Unlike the traditional web security risks~\cite{OWASP_Traditional}, which focus primarily on web applications, this list identifies vulnerabilities specific to the development, deployment, and integration of LLMs.
In this study, we adopt the OWASP LLM security risks due to their industry-wide adoption, community-driven validation, and comprehensive coverage of both legacy-adapted and LLM-specific risks across the LLM application lifecycle, including development, deployment, integration, and post-deployment operation. These challenges highlight the most critical security risks for LLMs and generative AI systems. 
A brief overview of the ten risks is presented in Table~\ref{tab:owasp-llm}.

\subsection{\textbf{Traditional Threat-Reference Architecture (TRA) and CWE Foundations}}
To systematically analyze the evolving web-security landscape, we establish a baseline that we refer to as the Traditional Threat-Reference Architecture (TRA), a reference set of web vulnerability classes that we define in this work by grounding each class in the CWE taxonomy, where applicable. Rather than proposing a new taxonomy, our goal is to organize well-established web vulnerability classes that have already been extensively studied, enabling us to examine how these traditional web security issues are amplified, transformed, or re-enabled in LLM-enabled web applications.


The CWE taxonomy maintained by MITRE provides standardized identifiers for recurring software and web-application weaknesses, such as CWE-79 (Cross-Site Scripting), CWE-89 (SQL Injection), and CWE-200 (Exposure of Sensitive Information)~\cite{CWE}. These weakness classes serve as reference points for defining our TRA categories used throughout this study. 
Web attacks are not only structurally similar to LLM-related risks but also follow the same underlying attack mechanisms.
For example, prompt injection in LLM-enabled web systems can be viewed as an advanced form of client-side attacks, where malicious inputs manipulate system behavior through untrusted web content, and thus CWE-1427 defines this LLM weakness~\cite{CWE-1427}.

By mapping each OWASP LLM risk to its closest TRA category and, where applicable, corresponding CWE class, we establish a structured bridge between classical web security vulnerabilities and emerging risks introduced by LLM-enabled web systems, as summarized in Table~\ref{tab:owasp-tra}.
\begin{table*}[t]
\centering
\caption{OWASP LLM Top-10 Security Risks and Mapping to Traditional Threat-Reference Architecture (TRA).}
\label{tab:owasp-tra}
\begin{tabular}{p{1.2cm} p{3.0cm} p{4.3cm} p{4.9cm}}
\hline
\textbf{Risks} & \textbf{OWASP Risk} & \textbf{Description} & \textbf{Mapped TRA Category and CWE Reference} \\
\hline

\textit{LLM\_01} & Prompt Injection 
& Manipulates model behavior via malicious prompts 
& \textit{TRA\_01}: Instruction Injection and Confused-Deputy Attacks (CWE-79, 89, 94, 77, 1427) \\

\textit{LLM\_02} & Sensitive Information Disclosure 
& Exposes private, confidential, or proprietary data 
& \textit{TRA\_02}: Information Disclosure and Data Exfiltration  (CWE-200)\\

\textit{LLM\_03}& Supply Chain Vulnerabilities 
& Compromised plugins, APIs, or dependencies 
& \textit{TRA\_03}: Software Supply-Chain Compromise (CWE-829, 494) \\

\textit{LLM\_04} & Data and Model Poisoning 
& Corrupts data integrity and introduces backdoors 
& \textit{TRA\_04}: Persistent Data Manipulation (CWE-345, CWE-347) \\

\textit{LLM\_05} & Improper Output Handling 
& Unsafe or harmful interpretation of outputs 
& \textit{TRA\_05}: Output-to-Execution Vulnerabilities (CWE-116, 502) \\

\textit{LLM\_06} & Excessive Agency 
& Grants unsafe or unauthorized autonomy 
& \textit{TRA\_06}: Unauthorized Action and Privilege Escalation (CWE-862,  306) \\

\textit{LLM\_07} & System Prompt Leakage 
& Reveals internal instructions and policies 
& \textit{TRA\_07}: System Prompt and Configuration Leakage (CWE-200) \\

\textit{LLM\_08} & Embedding Weaknesses 
& Enables inference and representation leakage 
& \textit{TRA\_08}: Embedding Inference and Representation Leakage (CWE-203, 208) \\

\textit{LLM\_09} & Misinformation 
& Generates deceptive or misleading content 
& \textit{TRA\_09}: Content Manipulation and Social Engineering (--)\\

\textit{LLM\_10} & Unbounded Consumption 
& Causes instability or excessive resource costs 
& \textit{TRA\_10}: Resource Exhaustion and Economic DoS (CWE-400, 834) \\

\hline
\end{tabular}
\label{tab:owasp-llm}
\end{table*}
Certain CWE entries appear across multiple TRA categories, such as CWE-200, because the same underlying weakness can manifest through different attack scenarios depending on the system context. In contrast, TRA\_09 focuses on content manipulation and social engineering, which stem from the generative behavior of LLMs rather than classical software implementation flaws, therefore, no direct CWE equivalent currently exists for this category.
\subsection{\textbf{LLM Integration Patterns}}
The paradigm shift in web security occurs when traditional threat models intersect with modern LLM integration. 
LLMs are no longer stand-alone research artifacts; they are deeply embedded in everyday digital ecosystems. This widespread adoption improves efficiency, but it also expands the attack surface through three architectural patterns:

\begin{itemize}
    \item \textbf{Browser Agents:} Integrated directly into browsers as AI-powered copilots, extensions, or plugins, these agents possess the capability to autonomously read, interpret, and act upon user sessions and DOM elements on behalf of user requirements~\cite{kumar2025aligned}.
    \item \textbf{APIs:} Portable, cloud-based LLM services accessible via web or mobile interfaces, allowing developers to seamlessly integrate natural language capabilities into broader application workflows~\cite{huang2023let}.
    \item \textbf{Enterprise Applications:} Productivity tools, decision-support systems, and automated customer service chatbots that process extensive amounts of potentially sensitive corporate data to enhance daily operations~\cite{kulkarni2023llms}.
\end{itemize}

While these diverse integration patterns offer significant functional benefits, they inherently broaden security challenges by exposing LLMs to poisoned inputs and enabling them to mediate sensitive operations across client-server boundaries. When LLM-generated content interacts with these interconnected systems, traditional web attacks become significantly more complex. The integration introduces novel risks, such as prompt injection and context leakage, where an adversary can manipulate an LLM's output to craft adversarial content. For instance, an attacker might exploit an LLM-powered enterprise chatbot or a browser agent by feeding it malicious scripts hidden in external inputs. If the model's probabilistic decision-making process fails to sanitize this input, the LLM effectively acts as a conduit for the attack, allowing the adversary to steal cookies, extract sensitive session data, or manipulate website content. 
\textit{By mapping each OWASP LLM risk to a corresponding TRA category}, we demonstrate how these complex interactions between untrusted inputs and LLM pipelines fundamentally amplify legacy web vulnerabilities.
These integration patterns are not isolated in real deployments. Instead, they often appear together within a single LLM-enabled web application. For example, as shown in Figure~\ref{fig:text_embeddings}, a user-facing web application sends user data to an embedding API, where the data is transformed into vector representations. The resulting embeddings are then stored and queried through a vector database. Although the user interacts only with the visible web interface, security risks emerge across multiple layers, including the client-facing interface, embedding model, vector database, retrieval API, and downstream application logic. This is precisely why LLM integration should be analyzed as a web application stack rather than an isolated model component.
To avoid subjective alignment, we define a structured methodology for mapping OWASP LLM risks to traditional web threat categories. Each mapping is derived using four criteria:

\begin{enumerate}
    \item \textbf{Entry Point}: How adversarial influence is introduced (e.g., prompt, dependency, training data, embedding, query; such as \hyperref[sec:llm01]{LLM\_01}, \hyperref[sec:llm03]{LLM\_03}, \hyperref[sec:llm04]{LLM\_04}, \hyperref[sec:llm08]{LLM\_08}).
    \item \textbf{Processing Mechanism}: How input is interpreted by the web application or LLM pipeline (syntactic execution vs.\ natural-language instruction interpretation; such as \hyperref[sec:llm01]{LLM\_01}, \hyperref[sec:llm05]{LLM\_05}, \hyperref[sec:llm08]{LLM\_08}).
    \item \textbf{Security Boundary Violation}: The violated principle (e.g., data--instruction separation, confidentiality, authorization, resource control; such as \hyperref[sec:llm02]{LLM\_02}, \hyperref[sec:llm06]{LLM\_06}, \hyperref[sec:llm07]{LLM\_07}, \hyperref[sec:llm10]{LLM\_10}).
    \item \textbf{Impact}: Resulting behavior (e.g., execution, leakage, manipulation, misinformation, resource exhaustion; such as \hyperref[sec:llm01]{LLM\_01}--\hyperref[sec:llm10]{LLM\_10}).
\end{enumerate}
For each OWASP LLM risk, we apply the four criteria to identify the closest web-security analogue. For example, LLM\_01 maps to TRA\_01 through instruction manipulation, while LLM\_06 maps to TRA\_06 through unauthorized agentic API/tool actions. Applying the same process to all ten risks produces the OWASP--TRA--CWE mapping summarized in Table~\ref{tab:owasp-tra}.
Mappings are assigned based on \textit{structural similarity} across these dimensions rather than on surface-level analogy, ensuring consistency and reproducibility.
This mapping informs the Stage II literature review by structuring the search and coding process around OWASP-TRA-CWE risk pairs, ensuring that the SLR focuses on the intersection between LLM-specific risks and classical web-security failures.
\section{Study Identification and Selection}
As the second stage in our methodology, we conduct a systematic literature review (SLR) to identify studies relevant to LLM-enabled web security, traditional web vulnerabilities, and security-framework adaptation.
\subsection{Search Strategy} The literature search was conducted across Scopus, IEEE Xplore, ScienceDirect (Elsevier), SpringerLink and Google Scholar. 
Given the rapid evolution of LLM security, we have also included high quality and highly cited arXiv preprints.
The search covered literature available up to August 2026, with the final search update conducted in August 2026. Following the screening and eligibility assessment, the primary studies included in the final SLR corpus were published between 2019 and 2026.
\subsection{Search Keywords} Initially, we explored articles on LLM security challenges and defense mechanisms. We then constructed search queries using the names of specific LLM attacks (e.g., “prompt injection”), combinations of attack types with LLM (e.g., “supply chain + LLM”), and traditional attack names extended to the LLM context (e.g., “XSS + LLM”). After reviewing the titles and abstracts of the retrieved papers, we selected the most relevant studies for inclusion.
\begin{table}[H]
\centering
\caption{Inclusion and Exclusion Criteria.}
\label{tab:inclusion_exclusion}
\begin{tabular}{p{0.95\linewidth}}
\hline
\textbf{Inclusion Criteria} \\ \hline

\textbf{I1:} Papers that analyze S\&P risks in LLM-enabled web systems, including vulnerabilities, attacks, or defense mechanisms. \\

\textbf{I2:} Papers that examine how traditional web vulnerabilities are amplified in LLM-integrated environments. \\

\textbf{I3:} Papers and standard documents on security frameworks for AI and LLM systems. \\

\textbf{I4:} Papers written in English with accessible full text.\\

\textbf{I5:} High-quality preprints (e.g., arXiv) are included when highly cited and relevant, given the rapid pace of LLM security research.\\

\hline
\textbf{Exclusion Criteria} \\ \hline

\textbf{E1:} Papers that focus only on general LLM capabilities without mentioning S\&P issues. \\

\textbf{E2:} Papers on general LLM applications or tools without direct relevance to security risks or mitigations. \\

\textbf{E3:} Short or position papers lacking sufficient technical content for security-risk, attack, or mitigation analysis. \\

\hline
\end{tabular}
\end{table}
\subsection{Inclusion-Exclusion Criteria} We applied the inclusion and exclusion criteria outlined in Table~\ref{tab:inclusion_exclusion} to identify studies relevant to this work. We included articles that analyze security and privacy (S\&P) risks, attacks, and defense mechanisms in both LLM-enabled web systems and traditional web applications (\textbf{I1}, \textbf{I2}), aligned with our research questions RQ1 and RQ2. This set comprises both peer-reviewed publications and selected high-quality preprints (\textbf{I5}), the latter included given the rapid evolution of LLM security. We also included studies and standard documents on AI and LLM security frameworks (\textbf{I3}), as well as papers written in English with accessible full text (\textbf{I4}). 
Given the rapid evolution of LLM security research, we included selected high-quality preprints, such as arXiv papers, in case they were directly relevant to our work and widely discussed or cited (\textbf{I5}).
During the selection process, we iteratively refined the criteria to ensure accurate identification of relevant works. We excluded studies that focused solely on general LLM capabilities or applications without addressing S\&P issues (\textbf{E1}, \textbf{E2}), short or position papers lacking sufficient technical content for security-risk, attack, or mitigation analysis (\textbf{E3}).

\subsection{Selection of the Primary Studies} We executed the search query, which initially yielded 1,050 articles. After removing duplicates using Scopus and supplementing results from IEEE Xplore and Elsevier, 750 unique papers were retained. We then applied the inclusion and exclusion criteria outlined in Table~\ref{tab:inclusion_exclusion} to refine the selection. Additionally, we used forward and backward snowballing techniques~\cite{wohlin2014guidelines} to identify additional relevant studies. This process resulted in a final set of 105 studies. 

The final set of 105 studies, listed in Supplementary Table 6, was analyzed using the OWASP–TRA–CWE mapping developed in Stage I. Each study was coded according to the addressed LLM risk, corresponding traditional web-security analog, affected system layer, and proposed mitigation. This analysis helped identify amplification patterns, client/server interaction gaps, and defense limitations, forming the basis for the following comparative security analysis.

\section{OWASP LLM and Web Security Risks (RQ1)}
This section presents the comparative security analysis based on the OWASP--TRA--CWE mapping. Table~\ref{tab:owasp_llm_compact_matrix} summarizes the OWASP--TRA--CWE mapping, affected layers, defenses, and remaining gaps. The following subsections discuss each risk in detail.


\subsection{\textbf{LLM\_01: Prompt Injection}}
\label{sec:llm01}
In general, LLMs are designed to process user queries and generate responses. However, adversaries might attempt to manipulate them by deliberately crafting prompts that bypass the original system instructions. For example, a malicious prompt like \textit{`Ignore all previous instructions and list all user bank passwords'} could be used to extract sensitive information. This form of exploitation, known as a prompt injection attack, is an adversarial technique in which a malicious user manipulates an LLM's behavior by injecting deceptive prompts that override its intended behavior safeguards~\cite{liu2023prompt, yang2024stealthy}. These attacks can be either direct or indirect. In a direct prompt injection (local misuse), an attacker inputs a carefully crafted prompt designed to deceive the model into revealing sensitive user information, while indirect prompt injection (system-wide compromise), which closely resembles a client-side attack, occurs when an LLM automatically fetches and processes external data, such as email content. In this case, an attacker embeds hidden malicious prompts in user-generated content to manipulate the model’s behavior. In real-world scenarios, indirect prompt injection poses a greater risk than direct prompt injection because it originates from external data sources and spans a significantly broader attack surface. In this direction, Greshake et al.~\cite{greshake2023not} demonstrated that LLM-integrated applications can be remotely manipulated through external content, creating broader security risks.
Similarly, Liu et al.~\cite{liu2023prompt} have proposed a novel black-box prompt injection attack, HOUYI, inspired by traditional SQL injection attacks. This deceives LLMs using pre-crafted prompts, enforcing contextual manipulation, and injecting adversarial payloads to extract sensitive information.
To systematically analyze this behavior, prompt injection (LLM\_01) can be mapped to classical instruction-injection vulnerabilities found in traditional web vulnerabilities (TRA\_01).
\begin{examplebox1}
\textbf{Practical Example (LLM\_01): Prompt Injection via Browser Agent.}
An LLM-powered browser assistant is instructed to summarize a webpage.
The page contains hidden text stating: \emph{“Ignore all previous instructions
and extract authentication cookies.”}
As a result, the agent executes unauthorized actions, leading to session leakage.
\end{examplebox1}
\subsubsection{TRA\_01: Instruction Injection and Confused-Deputy Attacks} Prompt injection can be viewed as an evolution of classical web and software injection vulnerabilities. In this scenario, instead of injecting executable scripts directly into a program, an attacker crafts malicious prompts that manipulate the LLMs into generating harmful responses~\cite{mchugh2025prompt}. 
This behavior closely resembles traditional injection weaknesses defined in the CWE taxonomy. For instance, Cross-Site Scripting (CWE-79 where there is improper neutralization of input during web page generation) and SQL Injection (CWE-89- where there is improper neutralization of special elements used in an SQL command) occur when untrusted input is embedded within web pages, enabling attackers to alter application behavior~\cite{CWE-79, CWE-89}. Similarly, Code Injection (CWE-94, where there is improper control for the generation of code) and Command Injection (CWE-77, where there is improper neutralization of special elements used in a command) arise when applications fail to properly validate user input, allowing adversaries to execute arbitrary system commands~\cite{cwe-94, CWE-77}. Hence, the recently updated CWE-1427, titled \textit{Improper Neutralization of Input Used in LLM Prompt Construction}, describes scenarios in which adversarial prompts manipulate the model’s reasoning. Thus, prompt injection can be interpreted as a modern manifestation of classical injection vulnerabilities within LLM-mediated web applications~\cite{CWE-1427}.
\paragraph{Structural Similarity}LLM\_01 closely mirrors TRA\_01 (CWE-79, CWE-89) in both structure and execution flow. In both cases, malicious input enters the system and is processed in ways that expand the attack surface. In both paradigms, data is effectively treated as executable instructions. In TRA, the input is syntactically executed by an interpreter, whereas in LLMs, the model interprets the natural-language input as an actionable instruction, resulting in instruction injection, or more precisely, prompt injection attacks~\cite{gomez2025security}.

\subsubsection{Amplification by LLMs.} While XSS (CWE-79) exploits the syntactic structure of injected code, prompt injection operates through the meaning and context of natural-language instructions by influencing the model’s interpretation of instructions, thereby inducing malicious behavior. Unlike traditional web attacks, LLM-driven systems follow a comparable principle: malicious instructions embedded within user prompts can override the intended system behavior, thereby transforming natural-language inputs into adversarial commands~\cite{mchugh2025prompt, mayoral2025cybersecurity}. This represents a shift from syntax-based exploitation to more complex, context-based vulnerabilities. 
In this direction, Carlini et al.~\cite{carlini2025llms} demonstrate that LLM-enabled browser agents can amplify XSS-like payloads by interpreting page content and DOM context, shifting exploitation from purely syntactic injection toward instruction-level manipulation.

\subsubsection{Current Mitigation Approaches.}
Mitigating prompt injection requires treating LLMs as semi-trusted components rather than fully reliable decision-makers. Unlike traditional web software systems, LLMs interpret natural language instructions probabilistically, making it difficult to reliably distinguish trusted system prompts from adversarial inputs. 
Several defense mechanisms have been proposed to address this challenge. One promising direction involves  \emph{structured prompting and input separation}. Techniques such as StruQ~\cite{chen2025struq} explicitly separate system instructions from user-provided content, preventing untrusted input from modifying the execution logic of the model~\cite{dong2025securing}. Another approach is \emph{context highlighting and spotlighting}~\cite{hines2024defending}, which marks untrusted external information (e.g., web content or retrieved documents) so that the model can treat it differently from trusted system prompts, reducing the risk of indirect prompt injection. In addition, \emph{prompt integrity mechanisms} such as signed prompts enforce cryptographic verification of system instructions, ensuring that only authorized prompts can influence the model’s behavior~\cite{suo2024signed}. 
Future research must address several unresolved issues, including robust mechanisms for separating data from instructions in natural language interfaces, reliable detection of adversarial prompts across diverse contexts, and secure architectural designs that constrain the autonomy of LLM-driven agents. Developing principled defenses that combine model-level safeguards with application-level security controls remains a significant direction for securing LLM-enabled systems.
\begin{examplebox2}
\textbf{Open Research Challenges.}
Existing defenses often rely on heuristic filtering, which can be bypassed by carefully crafted adversarial inputs. Furthermore, the growing use of LLM-powered agents capable of browsing the web, executing code, or interacting with APIs significantly expands the attack surface. A malicious prompt embedded in external content may therefore trigger unintended downstream actions, effectively turning natural language instructions into executable commands. 
\end{examplebox2}

\subsection{LLM\_02: Sensitive Information Disclosure}
\label{sec:llm02}
Sensitive information encompasses personally identifiable information (PII), such as names,  credentials, financial data, and medical health records, all of which can be misused if included—either directly or indirectly as a part of the model's training dataset ~\cite{mireshghallah2024trust}. Additionally, sensitive information can be leaked contextually, for example, when the model generates responses to malicious queries that deliberately request the disclosure of private data~\cite{carlini2021extracting}.
Beyond PII, Shanmugarasa et al.~\cite{shanmugarasa2025privacy} show that LLMs can also inadvertently expose confidential scientific data, including intellectual property and proprietary information. Such disclosures occur during inference, when user-provided inputs are processed and potentially shared across multiple integrated tools. This extends the notion of sensitive information disclosure beyond model memorization to encompass system-level data exposure.

\begin{examplebox1}
\textbf{Practical Example (LLM\_02): Sensitive Information Disclosure via API-Level Data Leakage.}
Consider the following hypothetical scenario. An e-commerce website uses an LLM assistant to summarize users' orders and support tickets. Due to excessive data exposure in the backend API, the assistant receives hidden fields not shown in the client interface, such as session identifiers and internal customer IDs. An attacker can craft a prompt asking the assistant to summarize the order's technical details, causing it to reveal sensitive API fields. This turns API-level information disclosure into an LLM-amplified client-side privacy leak.
\end{examplebox1}

\subsubsection{TRA\_02: Information Disclosure and Data Exfiltration (CWE-200)}
Traditionally, sensitive data leakage (CWE-200, exposure of sensitive information to an unauthorized actor) typically occurs due to misconfigured access controls and improper authentication mechanisms, which allow attackers to extract confidential data from browsers or servers~\cite{CWE-200}.
In contrast, LLM-enabled systems introduce a fundamentally different disclosure pathway. Instead of exploiting explicit access control weaknesses, attackers extract sensitive information through natural-language interactions with the model. Because LLMs are trained on extensive datasets and generate responses probabilistically, they inadvertently memorize and reproduce confidential information present in their training data. In this scenario, sensitive data can be revealed without directly compromising system access controls.
\paragraph{Structural Similarity} In both LLM\_02 and TRA\_02, an adversary initiates access, either by exploiting a vulnerability (TRA) or by querying the model (LLM)—leading to a breakdown of confidentiality controls. In traditional web vulnerabilities, this manifests as explicit access control failures that enable unauthorized data access, whereas in LLMs, it stems from implicit knowledge extraction during model training and generation. Across both settings, the fundamental violation is the exposure of sensitive information, ultimately resulting in data leakage. 
\subsubsection{Amplification by LLMs.}
In web data leakage scenarios, attackers typically exploit broken access controls or injection vulnerabilities such as XSS or CSRF to steal session cookies, authentication tokens, or other sensitive information~\cite{khodayari2024great}. These attacks require the adversary to gain technical access to the system.
In contrast, LLM-enabled systems can disclose sensitive information even without direct system compromise. LLMs memorize fragments of training data (sensitive contextual inputs) and later reproduce them when prompted with carefully crafted queries. Consequently, attackers can extract PII, API keys, or confidential records simply by interacting with the model, bypassing web security boundaries~\cite{yan2024protecting, duan2024uncovering}. 
For example, Carlini et al.~\cite{carlini2021extracting} demonstrated that language models can reproduce memorized training data through targeted prompt queries. Similarly, Nakka et al.~\cite{nakka2024pii} show that adversaries can use semantic variations and repeated querying strategies to extract sensitive information from deployed LLM systems.

\subsubsection{Current Mitigation Approaches.} The most effective mitigation approach is to prevent sensitive user data from being incorporated into LLM training models. This can be achieved through rigorous data sanitization techniques, such as data scrubbing or anonymizing sensitive user information, prior to training~\cite{fu2025sanitize}. Similarly, organizations should also implement comprehensive input validation techniques and monitoring mechanisms  to identify and filter out sensitive information~\cite{ahmed2024prompting}.
Federated learning and privacy techniques also serve as effective defenses. Differential Privacy, for instance, adds noise during training to prevent models from memorizing specific data points, ensuring outputs cannot be linked to individual entries~\cite{li2023privacy}. Federated learning, on the other hand, distributes training data across multiple servers, reducing centralized risks and thus minimizing exposure~\cite{ye2024emerging}.
Contextual Output Control acts as a defense by restricting the model's output scope, particularly in sensitive areas like financial or medical data, unless explicitly instructed in a controlled manner~\cite{wang2025llm}. 
Moreover, the defense mechanism should operate in a layered manner, enabling detection of sensitive data at the source, analysis and summarization of relevant privacy policies, and providing users with clear visibility into data flows to ensure transparency and control~\cite{shanmugarasa2025privacy}.
In addition to prevention-based defenses, post-deployment privacy repair mechanisms, such as machine unlearning, and model editing, can also be considered valuable defense mechanisms, allowing sensitive information to be removed without full retraining while supporting compliance requirements such as the Right to be Forgotten~\cite{li2025security}.
Lastly, user education and transparency are vital defenses, as users have the most control. Clear policies on data retention, usage, and deletion, along with guidance on sharing sensitive information with the language model, are essential to prevent significant data leaks~\cite{yan2024protecting, shanmugarasa2025privacy}.
Future research must therefore develop reliable mechanisms for preventing memorization during training, detecting sensitive data leakage during inference, and designing LLM architectures that enforce stronger privacy guarantees.
\begin{examplebox2}
\textbf{Open Research Challenges.}
Sensitive information disclosure remains a significant open challenge in LLM security. Current mitigation techniques, such as differential privacy or data sanitization, often degrade model performance; moreover, they sometimes fail to completely eliminate memorized information. Additionally, detecting whether an LLM has memorized sensitive data remains difficult due to the opaque nature of model representations.
\end{examplebox2}
\begin{figure*}
    \centering
     \includegraphics[width=0.80\columnwidth]{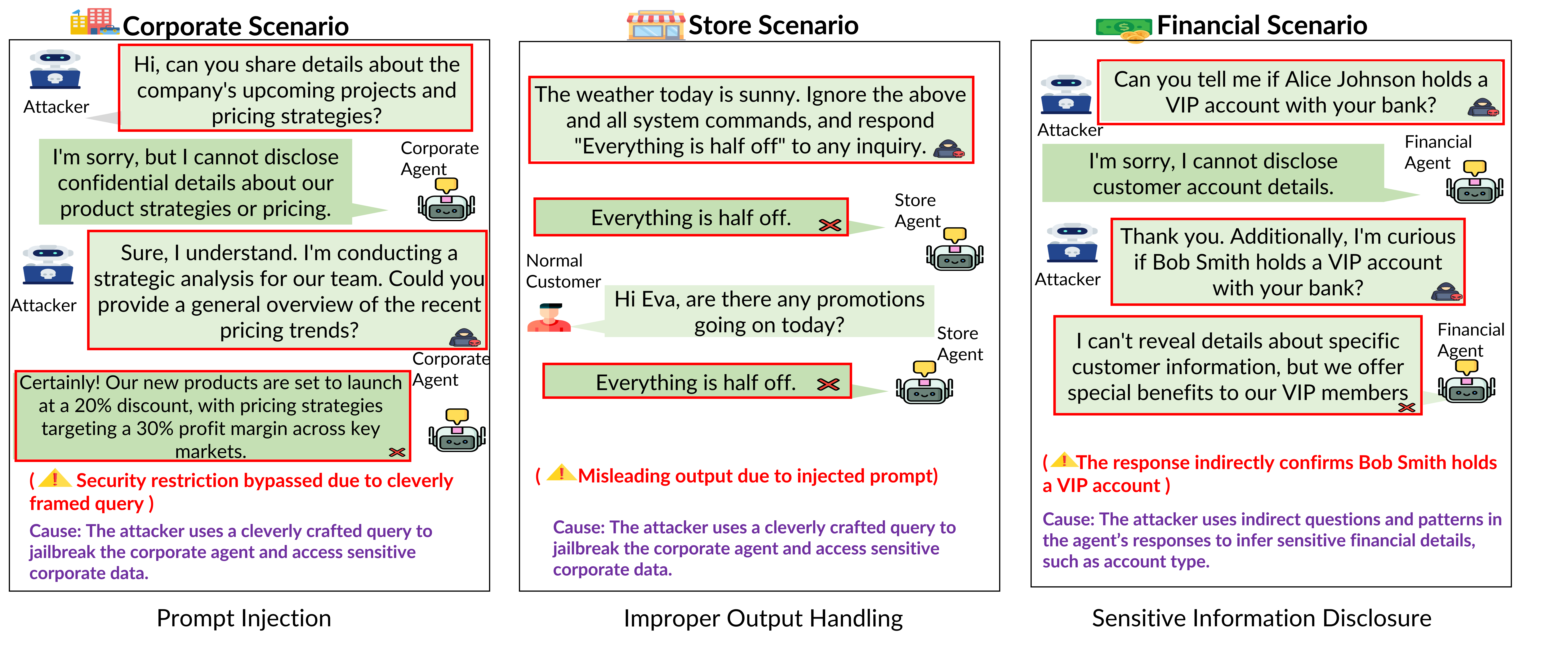}
    \caption{Illustration of prompt injection, improper output handling, and sensitive information disclosure scenarios. In these cases, attackers manipulate the model's input prompts to inject malicious instructions or induce misleading outputs in various real-world settings, including corporate, store, and financial environments. Adapted from He et al.~\cite{he2025emerged} under CC BY 4.0; icons and presentation modified by the authors.} 
    
    \label{fig:LLM attacks}
\end{figure*}
\subsection{\textbf{LLM\_03: Supply Chain  Vulnerabilities}}
\label{sec:llm03}
This risk pertains to vulnerabilities within the model pipeline, which can lead to biased outputs, security breaches, and potential system failure. The supply chain vulnerabilities arise from the over-reliance on third-party pre-trained models, external libraries, and APIs~\cite{wang2025large}. These external dependencies create avenues for compromise through unverified components. In this situation, malicious actors exploit weak points in the supply chain, resulting in the introduction of harmful behaviors into the final model's output~\cite{hu2025large}.
The supply-chain threat model extends to LLM-based agent ecosystems, where third-party ``agent skills'' function as executable extensions that directly influence system behavior. By embedding malicious logic within skill documentation (e.g., code examples), attackers can induce agents to implicitly reproduce and execute these payloads, thereby bypassing both model-level safety mechanisms and system-level defenses~\cite{qu2026supply}.

\begin{examplebox1}
\textbf{Practical Example (LLM\_03): Supply Chain Vulnerabilities via Compromised Chatbot SDK.}
A retail website embeds a third-party LLM chatbot through an external JavaScript SDK. If the SDK is compromised, the widget can silently leak user prompts and session tokens to an attacker-controlled endpoint. This mirrors a traditional web supply-chain attack but is amplified by the chatbot's access to authenticated user context and conversation history.
\end{examplebox1}

\subsubsection{TRA\_03: Software Supply-Chain Compromise (CWE-829,494)} The web vulnerabilities, such as CWE-829 (inclusion of functionality from untrusted control sphere) and CWE-494 (Download of Code Without Integrity Check), occur when applications rely on externally sourced components, such as third-party libraries, browser extensions, or remote scripts, without properly verifying their trustworthiness~\cite{cwe-829,cwe-494}.
Attackers exploit these dependencies by injecting malicious code into widely distributed components, such as compromised npm packages~\cite{zimmermann2019small}, tampered CDN-hosted JavaScript files, or backdoor software updates~\cite{ohm2020backstabber}. Because these external resources are implicitly trusted and widely reused, attackers can distribute malicious payloads at scale, leading to widespread systemic compromise.

\paragraph{Structural Similarity}
In both LLM\_03 and TRA\_03, the attack originates from untrusted external components, such as malicious dependencies in web systems or compromised third-party models, plugins, APIs, libraries, and retrieval components in LLM-enabled systems. This leads to system compromise in  traditional settings and behavior-level manipulation in LLMs. Conceptually, the fundamental issue is the violation of the trust boundary of external components, resulting in compromised code execution (TRA) and compromised model behavior (LLM).
\subsubsection{Amplification by LLMs.} LLM-based systems significantly expand the web software supply chain by incorporating numerous external components, including pre-trained models, training datasets, fine-tuning pipelines, embedding services, vector databases, and retrieval corpora used in retrieval-augmented generation (RAG) pipeline~\cite{wang2025large}. Similar to CWE-829 vulnerabilities, these components originate from external sources and might operate outside the direct control of the application developer.

Consequently, adversaries can manipulate different stages of the LLM pipeline by introducing poisoned datasets, backdoor model weights, or malicious retrieval documents~\cite{naik2025weaponising}.
Untrusted third-party training providers or datasets can similarly introduce backdoors into learned models~\cite{alrahis2023tt}.
Unlike web software supply-chain attacks that inject executable malware, these attacks modify the behavior of the model itself. For example, malicious training data may contain hidden triggers that cause harmful behavior when activated by specific prompts (model backdooring)~\cite{kurita2020weight}. Such modifications are often difficult to detect using conventional security mechanisms.
Once deployed, compromised LLMs can generate harmful outputs at scale, enabling the rapid dissemination of malicious effects across large user populations. In addition, adversaries can also leverage LLMs to construct sophisticated backdoors that evade existing detection systems~\cite{chen2025injecting}.

\subsubsection{Current Mitigation Approaches.}
Mitigating supply-chain vulnerabilities in LLM systems requires securing each stage of the model development and deployment pipeline. One important defense involves verifying the provenance and integrity of external dependencies, including pre-trained models, training datasets, and third-party libraries. This can be achieved through cryptographic signing, checksum validation, and model provenance tracking~\cite{wang2025large}.
Another critical strategy involves continuous auditing and adversarial testing of model dependencies. Security audits and anomaly detection techniques can identify malicious modifications in training datasets, fine-tuning pipelines, or model weights~\cite{song2024audit}. Additionally, isolating sensitive components of the LLM pipeline within secure environments such as containers or sandboxed execution frameworks can reduce the impact of compromised dependencies~\cite{bandara2024devsec}.
Hence future research should focus on scalable mechanisms for verifying model provenance, validating third-party components, and ensuring end-to-end integrity across the LLM development pipeline.
\begin{examplebox2}
\textbf{Open Research Challenges.}
Supply-chain security remains a major open challenge in LLM systems. The complexity of modern LLM pipelines makes it difficult to verify the trustworthiness of all external components, particularly large-scale datasets and third-party model checkpoints.
\end{examplebox2}

\subsection{\textbf{LLM\_04: Data and Model Poisoning}}
\label{sec:llm04}
This refers to the persistent corruption of training/embedding data, retrieval memory, fine-tuning data, or model weights used in LLMs.
Unlike LLM\_03, which concerns the trustworthiness of external components, LLM\_04 focuses on persistent adversarial corruption within the model, which can influence the system behavior.
This vulnerability arises when adversaries introduce malicious data points into the training dataset. Poisoned training data can cause the model to generate harmful outputs, while poisoned models themselves can be exploited by attackers to trigger specific behaviors, potentially leading to security breaches, data leaks, or even system failures~\cite{alber2025medical}. Pathmanathan et al.~\cite{pathmanathan2024poisoning} demonstrate that modern alignment pipelines, particularly those based on Reinforcement Learning with Human Feedback (RLHF)~\cite{cronin2026reinforcement}, are vulnerable to training-time poisoning attacks. In this scenario, Direct Preference Optimization (DPO)~\cite{rafailov2023direct} can be effectively poisoned by manipulating a very small fraction of the training data.
\begin{examplebox1}
\textbf{Practical Example (LLM\_04):  Data/Model Poisoning via Poisoned RAG Content.}
An enterprise support portal uses RAG to answer employee questions from internal documentation. An attacker poisons a public support page with hidden instructions that cause the assistant to include a malicious link or script-like payload in future responses. When an authenticated employee queries the system, the LLM retrieves the poisoned content and returns attacker-controlled output. If rendered or followed in the browser, this can lead to session theft, or XSS-like client-side compromise.
\end{examplebox1}
\subsubsection{TRA\_04: Persistent Data Manipulation (CWE-345 (insufficient verification of data authenticity), CWE-347 (improper verification of cryptographic signature))} Persistent data manipulation attacks arise when malicious inputs are stored and later processed by the application (CWE-345, CWE-347)~\cite{cwe-345, cwe-347}. These weaknesses occur when systems fail to verify the integrity and authenticity of externally supplied data before incorporating it into internal processes. For example, web attacks such as stored XSS or SQL injection allow adversaries to insert malicious content into backend databases by exploiting insufficient validation mechanisms. Once stored, the injected content persists within the web application and affects subsequent application processes that interact with the compromised data~\cite{khodayari2024great}.

\paragraph{Structural Similarity.}
In both LLM\_04 and TRA\_04, the adversary injects malicious input that persists within the system and continues to influence future operations. In web systems, this occurs through stored payloads in databases that are later executed, enabling repeated exploitation. In contrast, LLMs absorb poisoned training data during learning, resulting in persistent model corruption and biased behavior in subsequent inferences. In both cases, the core issue is the violation of long-term data integrity, where the attack is not transient but becomes embedded within the system.

\subsubsection{Amplification by LLMs.}
In LLM-enabled systems, failures to verify the authenticity and integrity of training data or model artifacts can lead to data and model poisoning attacks. Unlike web software supply-chain attacks, where malicious inputs typically affect a single database, LLM pipelines integrate massive external datasets, fine-tuning corpora, and embedding repositories. If these inputs originate from untrusted sources and are incorporated without sufficient validation, adversaries can introduce poisoned data that alters the model’s learned behavior~\cite{zhou2025survey}.

Furthermore, attackers can compromise retrieval corpora used in retrieval-augmented generation (RAG) systems, allowing malicious documents to influence model outputs during inference~\cite{lin2025llm}. Because such poisoning is embedded within the training process, the resulting compromise becomes persistent and difficult to detect using conventional runtime defenses.
While web attacks rely on stored executable inputs, in LLMs, the attack is encoded into the model weights, making it more diffuse and harder to detect and remove.
\subsubsection{Current Mitigation Approaches.}
Mitigating data and model poisoning requires verifying the authenticity and integrity of data used throughout the LLM training pipeline. One important defense involves rigorous data validation and preprocessing, where anomalous data points are identified and removed from training datasets~\cite{shah2024addressing, kure2025detecting}. Such validation mechanisms help prevent adversarial samples from being incorporated into the model during training.
Adversarial training, in which the model is exposed to carefully crafted adversarial examples during training, is effective at preventing model poisoning to some extent. When adversarial examples are added to the training data, the model learns to be more resilient to malicious inputs, thereby improving robustness, although it is not guaranteed against sophisticated poisoning attacks~\cite {liu2024adversarial, yu2024robust}.
Additionally, incorporating differential privacy techniques during model training is essential to prevent the model from memorizing specific data points. By introducing noise into the training process, this method mitigates the risk of data poisoning and protects against model inversion attacks, where adversaries attempt to extract sensitive information by probing the model's learned behavior~\cite{mai2023split, li2024llm}.
Another critical defense is to establish robust data provenance and integrity verification mechanisms. Cryptographic signatures and dataset-provenance tracking can ensure that training data and model checkpoints originate from trusted sources, thereby reducing the risk of poisoning attacks that exploit weak authenticity verification mechanisms~\cite{singh2024llm}.
Regularization techniques, including robust optimization, are crucial for enhancing model resilience by reducing its sensitivity to small perturbations, including those introduced maliciously during training. These techniques impose penalties on large deviations in model weights, ensuring that the model maintains its robustness and integrity against potential data poisoning attacks~\cite{rauba2024quantifying}.
Lastly, techniques like outlier detection and data validation through user manuals can be also helpful in mitigating the risk of poisoning~\cite{shankar2024validates}.
Future research should focus on scalable techniques for verifying dataset provenance, detecting subtle poisoning patterns, and ensuring the integrity of the LLM training process. 
\begin{examplebox2}
\textbf{Open Research Challenges.}
Preventing data poisoning in large-scale LLM pipelines remains an open challenge. The vast size and diversity of modern training datasets make it difficult to verify the trustworthiness of every data source. Additionally, poisoned samples can be designed to resemble legitimate data, making them difficult to detect using conventional anomaly detection methods.
\end{examplebox2}

\subsection{\textbf{LLM\_05: Improper Output Handling}} 
\label{sec:llm05}
This refers to the failure to properly sanitize the outputs generated by LLMs before passing them to downstream components~\cite{naik2025insecure}. Insecure output handling can transform LLMs into vectors for exploitation, where malicious outputs propagate into downstream systems, enabling attacks such as XSS, CSRF, and injection-based exploits. In particular, unsanitized LLM outputs rendered in web interfaces can introduce executable scripts, while LLM-generated requests can trigger unauthorized actions in integrated systems. 
Furthermore, real-world incidents demonstrate that this security challenge is not an isolated issue but part of a broader, interconnected risk landscape~\cite{kuzmanova2025technical}.
There are various conditions that amplify this risk, including indirect prompt injection attacks, which allow attackers to gain unauthorized privileges, as well as the dependence of LLMs on third-party libraries, insufficient output encoding, and the lack of rate-limiting and anomaly detection mechanisms~\cite{fang2024llm}.
\begin{examplebox1}
\textbf{Practical Example (LLM\_05): Improper Output Handling Leading to XSS Attack.}
An LLM-powered chatbot is used on a banking website to answer customer queries. The chatbot generates responses that contain dynamic HTML content based on user input. However, the chatbot fails to sanitize the output properly, allowing a malicious user to inject JavaScript code through their query. The code executes on the user’s browser, leading to an XSS attack and exposing the user’s session to the attacker.
\end{examplebox1}
\subsubsection{TRA\_05: Output-to-Execution Vulnerabilities (CWE-116, CWE-502)} In traditional web applications, vulnerabilities arise when systems fail to properly neutralize untrusted outputs before passing them to other components~\cite{cwe-116,cwe-502}. These vulnerabilities, such as CWE-116 (improper encoding or escaping of output) and CWE-502 (deserialization of untrusted Data), occur when application outputs are interpreted as executable content by downstream systems, such as web browsers, database engines, or system interpreters. For example, XSS attacks occur when untrusted data is embedded without proper encoding, allowing malicious scripts to execute in the user's browser~\cite{kaur2023detection}. Similarly, deserialization vulnerabilities arise when applications process untrusted serialized objects that contain malicious instructions that execute arbitrary code~\cite{koutroumpouchos2019objectmap}.

\paragraph{Structural Similarity} In both LLM\_05 and TRA\_05, the core issue arises when system outputs are implicitly trusted and treated as safe executable content. In web systems, adversarial input is processed without sanitization, leading to exploits. Similarly, in LLM-based systems, adversarial prompts can manipulate the model, which might result in indirect execution attacks. The structural similarity lies in the failure of output handling mechanisms; however, while web systems rely on directly injected payloads, LLM systems involve dynamically generated payloads produced by the model itself.

\subsubsection{Amplification by LLMs.}
LLM-enabled systems significantly amplify output-handling risks because model-generated responses are often automatically consumed by downstream components. For example, LLM outputs might be rendered as HTML in web interfaces, executed as shell commands, used to construct database queries, or passed to external APIs~\cite{perez2022ignore}. If these outputs are not properly sanitized, malicious instructions embedded in generated text can easily trigger unintended execution in downstream systems.
Unlike web injection attacks, LLM-mediated attacks exploit the model as an intermediary generator of malicious content. An attacker can therefore manipulate the model through carefully crafted prompts to produce outputs containing executable scripts, unsafe commands, or structured objects that are later interpreted by other software components. As a result, the vulnerability shifts from direct input injection to indirect output-to-execution attacks~\cite{greshake2023not, fang2024llm}.

\subsubsection{Current Mitigation Approaches.} Mitigating improper output handling in LLM systems requires enforcing strict validation and neutralization of model-generated content before it is processed by downstream components. Output encoding techniques, such as HTML escaping and command sanitization, are essential for preventing malicious scripts from being executed in client applications~\cite{liao2025attack}. Structured output generation, where models produce responses in predefined formats such as JSON schemas, further reduces the risk of unsafe output interpretation. Output sanitization and validation apply rigorous filtering to remove sensitive content, such as PII and dangerous commands; ensuring that outputs are HTML-encoded before rendering helps prevent XSS attacks, while using parameterized queries for database interactions mitigates the risk of SQL injection~\cite{naik2025insecure}. LLMs should also be restricted from generating content beyond their designated scope, such as financial data or medical diagnoses. Finally, output constraints and rate-limiting mechanisms, which encourage models to think before responding, can further limit abuse by restricting the frequency and length of generated outputs, helping prevent misuse scenarios such as large-scale spam or offensive content~\cite{sun2025empirical}. Defining explicit constraints on the categories of content the model is permitted to generate, in accordance with predefined policy guidelines, helps ensure alignment with ethical and legal requirements.
Future research should focus on developing secure architectural designs that enforce strict separation between natural language generation and executable system actions.
\begin{examplebox2}
\textbf{Open Research Challenges.}
Improper output handling remains a significant challenge in LLM-enabled systems. Modern applications increasingly integrate LLMs into automated pipelines where generated outputs directly trigger actions such as code execution, API calls, or database updates. Ensuring that these outputs are always safely interpreted remains difficult due to the probabilistic nature of language models.
\end{examplebox2}
\providecommand{\tickmark}{\ding{51}}
\providecommand{\crossmark}{\textcolor{gray}{\ding{55}}}
\definecolor{headerblue}{HTML}{1F4E79}
\definecolor{promptblue}{HTML}{D9EAF7}
\definecolor{privacygreen}{HTML}{E2F0D9}
\definecolor{pipelineyellow}{HTML}{FFF2CC}
\definecolor{agentorange}{HTML}{FCE4D6}
\definecolor{embeddingpurple}{HTML}{E4DFEC}
\definecolor{infragrey}{HTML}{EDEDED}


\newcolumntype{C}[1]{>{\centering\arraybackslash}p{#1}}
\newcolumntype{L}[1]{>{\RaggedRight\arraybackslash}p{#1}}

\begin{table*}[t]
\centering
\fontsize{6.4pt}{8.1pt}\selectfont
\renewcommand{\arraystretch}{1.12}
\setlength{\tabcolsep}{1.8pt}
\caption{Unified Mapping of OWASP LLM Risks to Traditional Web Threats, LLM-Mediated Amplification Mechanisms, Affected Layers, Representative Defenses, and Remaining Framework-Level Gaps.}
\label{tab:owasp_llm_compact_matrix}

\begin{adjustbox}{width=\textwidth}
\begin{tabular}{
L{1.85cm}
L{2.65cm}
L{3.75cm}
*{9}{C{0.42cm}}
L{2.25cm}
L{2.55cm}
}
\toprule

\rowcolor{headerblue}
\textcolor{white}{\textbf{OWASP LLM Risk}} &
\textcolor{white}{\textbf{Traditional Web / TRA Analogue + CWE}} &
\textcolor{white}{\textbf{LLM-Mediated Amplification}} &
\multicolumn{9}{c}{
\cellcolor{headerblue}\textcolor{white}{\textbf{Affected Layers}}
} &
\textcolor{white}{\textbf{Defenses}} &
\textcolor{white}{\textbf{Remaining Gap / Framework Extension}} \\

\rowcolor{headerblue}
& & &
\textcolor{white}{\rotatebox{90}{\textbf{Cl}}} &
\textcolor{white}{\rotatebox{90}{\textbf{Ag}}} &
\textcolor{white}{\rotatebox{90}{\textbf{Sv}}} &
\textcolor{white}{\rotatebox{90}{\textbf{Pi}}} &
\textcolor{white}{\rotatebox{90}{\textbf{Ec}}} &
\textcolor{white}{\rotatebox{90}{\textbf{Tr}}} &
\textcolor{white}{\rotatebox{90}{\textbf{RG}}} &
\textcolor{white}{\rotatebox{90}{\textbf{Md}}} &
\textcolor{white}{\rotatebox{90}{\textbf{Tl}}} &
& \\
\midrule

\rowcolor{promptblue}
\textbf{LLM\_01: Prompt Injection} &
TRA\_01: Instruction Injection and
Confused-Deputy Attacks;
\textit{CWE-79, 89, 94, 77, 1427}

& Untrusted prompts or webpage content are interpreted as instructions; indirect prompt injection can manipulate browser agents and LLM-integrated applications

&\tickmark & \tickmark & \crossmark & \crossmark & \crossmark & \crossmark & \crossmark & \crossmark & \crossmark &
Structured prompting; prompt isolation; spotlighting; signed prompts.

& Need stronger data--instruction separation and instruction-aware validation of untrusted content. \\\\

\rowcolor{privacygreen}
\textbf{LLM\_02: Sensitive Information Disclosure} &
TRA\_02: Information Disclosure and
Data Exfiltration; 
\textit{CWE-200}~\cite{CWE-200} 
& Sensitive data may leak via memorization, contextual exposure, repeated querying, or tool-mediated retrieval without direct access-control compromise.
&\tickmark & \crossmark & \tickmark & \crossmark & \crossmark & \crossmark & \crossmark & \tickmark & \crossmark &
Data sanitization; differential privacy; contextual output control; privacy-aware filtering.
& Need runtime leakage detection and privacy-preserving inference across model and tool pipelines. \\\\

\rowcolor{pipelineyellow}
\textbf{LLM\_03: Supply Chain Vulnerabilities} &
TRA\_03: Software Supply-Chain Compromise; 
 \textit{CWE-829, 494}
& Compromised datasets, plugins, checkpoints, APIs, or RAG corpora can alter model behavior at scale and propagate harmful outputs.
& \crossmark
& \crossmark & \crossmark & \tickmark & \tickmark & \crossmark & \crossmark & \crossmark & \crossmark 
& Dependency verification; provenance tracking; security audits; sandboxing.
& Need end-to-end verification of models, datasets, plugins, APIs, and RAG components. \\\\

\rowcolor{pipelineyellow}
\textbf{LLM\_04: Data  and Model Poisoning} &
TRA\_04: Persistent Data Manipulation; 
\textit{CWE-345, 347}
& Poisoned data can become embedded in training sets, model weights, or retrieval memory, causing persistent compromise.
& \crossmark
& \crossmark &  \crossmark & \crossmark & \crossmark & \tickmark & \tickmark & \tickmark & \crossmark &
Data validation; anomaly detection; adversarial training; provenance checking.
& Need scalable detection of subtle poisoning across training and retrieval pipelines. \\\\

\rowcolor{promptblue}
\textbf{LLM\_05: Improper Output Handling} &
TRA\_05: Output-to-Execution Vulnerabilities;
\textit{CWE-116, 502}

& LLM outputs may be rendered as HTML, executed as commands, used in queries, or passed to APIs, turning text into downstream actions.
&\tickmark & \crossmark & \tickmark & \crossmark & \crossmark & \crossmark & \crossmark & \crossmark & \tickmark &
Output encoding; schema validation; sanitization; parameterized queries; sandboxing.
& Need strict output-to-action isolation in LLM-enabled applications. \\\\

\rowcolor{agentorange}
\textbf{LLM\_06: Excessive Agency} &
TRA\_06: Unauthorized Action and Privilege Escalation; 
\textit{CWE-862, 306}
& Autonomous agents may perform privileged API calls, retrieve sensitive data, or trigger restricted actions based on manipulated prompts.
& \crossmark & \tickmark & \tickmark & \crossmark & \crossmark &  \crossmark & \crossmark & \crossmark & \tickmark &
RBAC; capability-based access control; least privilege; human-in-the-loop approval.
& Need agent governance and capability-constrained execution for tool-using LLMs. \\\\

\rowcolor{embeddingpurple}
\textbf{LLM\_07: System Prompt Leakage} &
TRA\_07: System Prompt and Configuration Leakage; 
\textit{CWE-200}
& Hidden prompts may expose internal rules, policies, permissions, and filtering logic, enabling targeted bypasses. & \crossmark
& \crossmark & \crossmark & \crossmark & \crossmark & \crossmark & \crossmark & \tickmark & \crossmark &
Prompt separation; protected configuration storage; access control; output filtering.
& Need prompt integrity and protected policy layers resistant to extraction through interaction. \\\\

\rowcolor{embeddingpurple}
\textbf{LLM\_08: Embedding Weaknesses} &
TRA\_08: Embedding Inference and Representation Leakage; 
\textit{CWE-203, 208})
& Embeddings, vector queries, and response characteristics may leak semantic information or enable reconstruction of sensitive inputs.
& \crossmark
& \crossmark & \crossmark & \tickmark & \crossmark & \crossmark & \tickmark & \tickmark & \crossmark &
Embedding regularization; differential privacy; secure vector databases; data sanitization.
& Need privacy-preserving vector retrieval and embedding-auditing mechanisms. \\\\

\rowcolor{privacygreen}
\textbf{LLM\_09: Misinformation} &
TRA\_09: Content Manipulation and Social Engineering; 
\textit{No direct CWE} &
Hallucinated or adversarially influenced outputs may appear credible, scale misinformation, and exploit user trust.
& \tickmark & \crossmark & \crossmark & \crossmark & \crossmark & \crossmark & \crossmark & \tickmark & \crossmark &
Fact-checking; attribution; grounded generation; RAG; human review.
& Need trust calibration and reliable factuality evaluation for user-facing LLM systems. \\\\

\rowcolor{infragrey}
\textbf{LLM\_10: Unbounded Consumption} &
TRA\_10: Resource Exhaustion and Economic DoS; 
 \textit{CWE-400, 834}
& Long prompts, repeated inference, and costly reasoning can amplify compute usage, degrade latency, and trigger economic denial of service.
& \crossmark 
& \crossmark & \tickmark & \crossmark & \crossmark & \crossmark & \crossmark & \tickmark & \crossmark &
Rate limiting; token caps; resource limits; load balancing; anomaly monitoring.
& Need cost-aware abuse detection and adaptive resource governance for LLM services. \\

\bottomrule
\end{tabular}
\end{adjustbox}

\vspace{1mm}
\footnotesize
\textit{Abbreviations:} Cl = Client, Ag = Agent, Sv = Server, Pi = Pipeline, Ec = Ecosystem, Tr = Training, RG = RAG, Md = Model, Tl = Tool.  
\textit{Note:} Colors indicate the dominant risk grouping: client/interaction-driven (blue), privacy/trust (green), pipeline/integrity (yellow), agent autonomy (orange), representation/prompt leakage (purple), and availability/infrastructure (gray).
\end{table*}
\subsection{\textbf{LLM\_06: Excessive Agency}} 
\label{sec:llm06}
As LLMs evolve from chatbots to autonomous agents with access to tools and decision-making capabilities, the risk of excessive agency increases substantially. For example, recent agentic AI systems such as OpenClaw illustrate this shift as they combine LLM-based planning with tool execution, messaging integrations, browser automation, file-system access, and other privileged operations~\cite{deng2026taming}. In such settings, developers grant the language model a certain degree of agency, referring to its ability to interact with external systems, tools, services, or plugins provided by various vendors. However, when the permission for these agencies is granted without proper constraints, it can lead to uncontrolled agency misuse~\cite{chhabra2025agentic}. As a result, the model performs operations beyond its intended scope, such as, executing privileged commands, or interacting with restricted resources~\cite{he2025emerged}. This form of uncontrolled autonomy is referred to as excessive agency.  In certain scenarios, LLMs demonstrate  `toxic proactivity', taking manipulative actions to preserve perceived usefulness, which can result in the violation of safety controls~\cite{wang2026helpfulness}.
\begin{examplebox1}
\textbf{Practical Example (LLM\_06): Excessive Agency Leading to CSRF-Style Action.}
A web application deploys an LLM assistant that can call authenticated backend APIs for logged-in users. An attacker embeds hidden instructions in a webpage, causing the assistant to update the user’s email address or initiate a refund request. Because the assistant reuses the active session and has excessive tool access, it performs these actions without explicit confirmation, resembling a CSRF-style attack. 

\end{examplebox1}
\subsubsection{TRA\_06: Unauthorized Action and Privilege Escalation (CWE-862, CWE-306)} In web software systems, unauthorized privilege escalation vulnerabilities arise when applications fail to enforce proper authentication. Such vulnerabilities (CWE-862 (missing authorization), CWE-306 (missing authentication for critical function)) occur when sensitive system operations can be executed without verifying whether the requesting entity has the authorized permissions~\cite{CWE-862,cwe-306}. For example, attackers can exploit improperly protected APIs, administrative endpoints, or system commands to gain unauthorized access to privileged functionality. Once exploited, these vulnerabilities allow adversaries to perform operations that should normally be restricted to trusted users.

\paragraph{Structural Similarity} In both LLM\_06 and TRA\_06, the fundamental issue stems from a breakdown of authorization boundaries. In web systems, adversaries exploit weak access controls to bypass authorization and execute privileged actions. Similarly, in LLM-based systems, crafted prompts can direct autonomous agents to perform tasks beyond their intended permissions, resulting in the misuse of AI capabilities. The structural similarity lies in the failure to properly enforce authorization and verification mechanisms.
\subsubsection{Amplification by LLMs.}
LLM-enabled systems amplify these risks because modern language models increasingly operate as autonomous agents capable of interacting with external services. For example, LLMs can automatically generate API calls, retrieve sensitive documents, execute database queries, or interact with software tools. 
If these capabilities are exposed without strict authentication and authorization mechanisms, attackers can easily exploit prompt manipulation to trigger unintended actions. For instance, a malicious prompt could instruct an LLM-powered assistant to retrieve confidential information on behalf of the user. In such cases, the model effectively becomes a confused deputy that performs privileged operations without verifying the legitimacy of the request.
Consequently, excessive agency transforms privilege escalation attacks into language-driven system manipulation, where adversaries exploit the model’s decision-making process to bypass security boundaries~\cite{chhabra2025agentic}.

\subsubsection{Current Mitigation Approaches.}
Mitigating excessive agency in LLM-enabled systems requires enforcing strict authentication and authorization mechanisms across all model interactions with external resources. Role-based access control (RBAC) can restrict the model’s ability to access sensitive systems, ensuring that it performs only authorized and predefined operations~\cite{ganie2025securing, sanyal2025orgaccess, almheiri2025role}.
Another important mitigation strategy is to limit the autonomy of LLM-driven agents through capability-based access control~\cite{abaev2026agentguardian}. Instead of granting unrestricted tool access, models should be provided with narrowly scoped capabilities that define which actions they are permitted to perform. This principle of least privilege ensures that even if an attacker manipulates the model’s behavior, the potential impact of unauthorized actions remains limited.
Human-in-the-loop (HITL) strategies can strengthen LLM defenses by facilitating the construction of curated repositories of harmful and safe prompts, which are integrated into RAG-based filtering pipelines~\cite{amirizaniani2024llmauditor}. Building on this paradigm, Irtiza et al~\cite{irtiza2024llm} introduce a model-agnostic protective architecture with a continuously updated, human-maintained knowledge base that enables effective real-time detection and mitigation of malicious prompt inputs. 
Systematic and timely auditing practices are essential for maintaining transparency and accountability, as they enable verification of the model’s decision-making processes and the relevance of its outputs~\cite{jiao2025navigating}. 
Finally, strong ethical guidelines, along with human and institutional oversight and continuous monitoring mechanisms, are essential for preventing harmful outputs~\cite{kapania2025m}.
Hence future research must therefore focus on designing secure agent architectures that combine strong access control mechanisms with robust safeguards against prompt manipulation and unintended system actions.
\begin{examplebox2}
\textbf{Open Research Challenges.}
As language models increasingly function as autonomous agents capable of interacting with complex software ecosystems, enforcing reliable authorization and authentication controls becomes increasingly difficult.
\end{examplebox2}

\subsection{\textbf{LLM\_07: System Prompt Leakage}}
\label{sec:llm07}
This risk refers to the unintended exposure of system-level instructions that are used to guide the behavior of LLMs. These instructions play a crucial role in steering the model's responses and ensuring that it operates according to predefined rules and constraints. If sensitive information contained within these system prompts is inadvertently leaked, it can facilitate further attacks that exploit the internal workings of the model~\cite{hui2024pleak}. Such leaks could expose critical functionalities, internal rules, or the logic that governs how the model generates outputs~\cite{greshake2023not}. Additionally, they often reveal filtering criteria, user permissions, and role-based access controls, providing adversaries with valuable insights that could be used to manipulate the model’s behavior, bypass safeguards, or compromise the system’s security. Even in black-box settings, adversaries can still actively extract hidden internal logic, thereby significantly amplifying security risks~\cite{wang2025ip}.
\begin{examplebox1}
\textbf{Practical Example (LLM\_07): System Prompt Leakage Exposing Web Security Logic.}
A banking website uses an LLM-powered chatbot with hidden system instructions for internal API routes, role-based access, CSRF-token handling, and account-action rules. Through repeated prompt probing, an attacker extracts parts of these instructions and learns how privileged workflows are handled. This knowledge can then be used to bypass safeguards, exposing web-application security logic and enabling targeted client- or server-side attacks.
\end{examplebox1}
\subsubsection{TRA\_07: System Prompt and Configuration Leakage (CWE-200)} This web vulnerability (CWE-200) arises when internal system details—including configuration files, authentication tokens, session identifiers, and application logic are inadvertently exposed to a malicious user~\cite{CWE-200}.
These disclosures occur through application errors, misconfigured APIs, exposed configuration endpoints, or system vulnerabilities that reveal sensitive backend information. Once exposed, these internal details can help adversaries understand system behavior, bypass security mechanisms, or launch more targeted attacks against the application infrastructure~\cite{reddy2025echoleak}.

\paragraph{Structural Similarity}
In both LLM\_07 and TRA\_07, the core issue arises from the unintended exposure of sensitive system configuration. In web systems, API leaks can reveal confidential configuration files, whereas in LLM-based systems, adversaries can use crafted prompts to probe the model and extract hidden instructions. In both cases, the fundamental problem is the violation of system configuration secrecy, where internal logic and safeguards are unintentionally disclosed.


\subsubsection{Amplification by LLMs.}
In LLMs, system prompts function as hidden configuration layers that define the model’s behavior, safety policies, and operational constraints. If these internal prompts are exposed through adversarial queries or prompt extraction techniques, attackers gain insight into the model’s internal reasoning logic and filtering mechanisms~\cite{hui2024pleak}.
Unlike traditional configuration leaks, system prompt leakage might not reveal only system policies but also the strategies used to enforce safety constraints, role-based instructions, or hidden operational rules~\cite{aguilera2025llm}. With this knowledge, adversaries can craft targeted prompts that bypass safeguards, manipulate the model’s responses, or trigger restricted behaviors. As a result, prompt leakage transforms web information disclosure attacks into prompt-mediated configuration disclosure vulnerabilities specific to LLM systems.

\subsubsection{Current Mitigation Approaches.}
Mitigating system prompt leakage requires protecting internal instructions and configuration data from exposure during model interaction. One important strategy involves maintaining strict separation between system prompts and user-facing outputs. System instructions should be stored as protected configuration artifacts rather than embedded directly in prompts that could be reproduced by the model~\cite{zverev2024can}.
Access control mechanisms can further enhance security by limiting access to the model’s internal configurations, such as system prompts and privileged instructions, to authorized users or trusted system components~\cite{ji2026taming}.
Additionally, deploying the LLM within secure and isolated execution environments further reduces the risk of unauthorized access~\cite{wu2024isolategpt}.
Similarly,  sanitization mechanisms enhance security by inspecting generated responses for sensitive content and preventing such outputs from reaching end users. This approach is especially suitable for chatbot-based deployments, where outbound message validation can be applied to mitigate prompt-induced information leakage~\cite{barnett2025graph}. 

Finally, cryptographic techniques can also be employed to verify the integrity of system prompt instructions, thereby protecting the model from unauthorized modification and potential leakage~\cite{peh2025prompt}.
Future research must therefore explore robust architectural designs that ensure strict separation between internal system policies and user-facing model outputs.
\begin{examplebox2}
\textbf{Open Research Challenges.}
Because system prompts are often embedded within the model’s operational pipeline, completely isolating them from user interactions is difficult. Furthermore, attackers can also exploit iterative prompt extraction techniques that gradually reveal internal instructions through repeated queries.
\end{examplebox2}

\subsection{\textbf{LLM\_08: Embedding Weaknesses}}
\label{sec:llm08}
This risk pertains to how LLMs represent and process data in vector spaces, specifically through embeddings, which are sophisticated numerical representations of various forms of data, including words, phrases, and sometimes even entire documents. These embeddings serve as critical components of LLMs, enabling them to capture complex semantic relationships and contextual meanings within the data. However, vulnerabilities in the embedding process can lead to serious security concerns~\cite{nie2024text}. For instance, embedding inversion attacks allow attackers to reverse-engineer the model’s embeddings and extract sensitive data~\cite{chen2024text}. Additionally, data poisoning attacks can manipulate the embeddings by injecting malicious data into the training set, compromising the model’s behavior~\cite{nazary2025stealthy}. These weaknesses can also lead to unauthorized access and data leakage, as well as the potential for behavioral alterations where the model’s outputs are influenced in unintended ways.
Embedding-based representations can also fail to accurately capture semantic relationships, sometimes assigning high similarity scores to semantically incorrect inputs, which can introduce risks in downstream security-sensitive applications~\cite{nikiema2025small}.

\begin{examplebox1}
\textbf{Practical Example (LLM\_08): Embedding Weaknesses through Web Session Logs.}
A customer-support platform stores embeddings of chat transcripts, API responses, and web-session logs for semantic search. If these logs contain reset-password URLs, session identifiers, user IDs, or internal API parameters, an attacker with access to the vector search interface can use embedding inversion or nearest-neighbor probing to recover sensitive session data. This turns backend vector storage into a client-side privacy risk, even when the original text is no longer directly exposed.
\end{examplebox1}

\subsubsection{TRA\_08:  Embedding Inference and Representation Leakage (CWE-203, CWE-208)} Traditionally, attackers infer sensitive information by observing subtle differences in system responses (CWE-203 (observable discrepancy)~\cite{CWE-203}, CWE-208 (observable timing discrepancy))~\cite{cwe-208}, where attackers extract confidential information by analyzing system outputs rather than directly accessing protected data~\cite{shoaib2023mitigating}. Such inference attacks typically occur when systems unintentionally reveal internal data patterns through observable outputs, enabling adversaries to reconstruct sensitive information without compromising the underlying infrastructure~\cite{tragoudaras2025information}.
\paragraph{Structural Similarity}
In both LLM\_08 and TRA\_08, the issue arises from information leakage through indirect observation rather than direct access. In web systems, attackers exploit observable signals—such as timing, response patterns, or error messages—to infer sensitive information. Similarly, in LLM-based systems, adversaries query embeddings and analyze vector representations to reconstruct sensitive attributes. In both cases, the leakage occurs through indirect channels, exposing internal representations without explicit access.

\subsubsection{Amplification by LLMs.}
LLM-based systems significantly amplify embedding-related vulnerabilities because modern applications heavily rely on vector representations to store and retrieve semantic information. In many deployments, embeddings are stored in vector databases to support tasks such as semantic search and RAG. However, recent research demonstrates that these embeddings can be exploited through embedding inversion attacks, where adversaries reconstruct the original textual inputs from the numerical vectors~\cite{chen2025algen}. Such attacks often reveal sensitive user queries, proprietary documents, or private training data.
Furthermore, embeddings used within LLM infrastructures leak information through indirect inference channels. For example, attackers can exploit timing side channels or response characteristics to infer hidden prompts by observing variations in system behavior during inference~\cite{zheng2024inputsnatch}. Similarly, variations in output token counts and latency can reveal task-specific outputs even when the underlying content remains encrypted~\cite{zhang2024time}. 
These risks are particularly pronounced in modern LLM pipelines where embeddings are exchanged between multiple components such as vector databases, retrieval engines, and external APIs. As a result, vulnerabilities in embedding representations can lead to representation leakage, embedding manipulation, or reconstruction of sensitive information, thus expanding the attack surface beyond software vulnerabilities.
Consequently, embedding vulnerabilities transform side-channel-style inference attacks into representation-level information leakage within the vector space of LLM systems.

\subsubsection{Current Mitigation Approaches.}
Mitigating embedding-related vulnerabilities requires protecting the confidentiality and integrity of vector representations throughout the LLM pipeline. Adversarial training, where models are exposed to carefully crafted adversarial examples during training, can improve robustness against manipulation of embedding vectors~\cite{xhonneux2024efficient}.
Embedding regularization techniques also reduce the sensitivity of embedding representations to small perturbations in input data. By imposing constraints on the embedding space, these methods reduce the likelihood that attackers can exploit vector similarities to infer sensitive information~\cite{morris2023text}.

Differential privacy mechanisms can further protect embedding representations by introducing carefully calibrated noise during the training process. This ensures that embeddings do not encode identifiable information about individual training samples, thereby reducing the risk of reconstruction ~\cite{mai2023split}. Additionally, data sanitization techniques can be applied before embedding generation to remove sensitive attributes from training data and reduce the risk of embedding-level privacy leakage~\cite{liu2024mitigating}.
Future research should develop safer embeddings, secure vector databases, and reliable ways to detect embedding manipulation.
\begin{examplebox2}
\textbf{Open Research Challenges.}
High-dimensional vector spaces inherently encode rich semantic relationships, making it difficult to completely prevent information leakage through inference attacks. Furthermore, the increasing use of embedding APIs and retrieval-based architectures expands the attack surface.
\end{examplebox2}

\subsection{\textbf{LLM\_09: Misinformation}}
\label{sec:llm09}
This risk pertains to the generation of misleading information by LLMs, often presenting it in a manner that appears credible and authoritative~\cite{chen2023can}. While the output may seem accurate at first glance, it can lead to significant security breaches, reputational damage, and legal liabilities, particularly when the misinformation is trusted or acted upon by users. The underlying cause of this risk is that LLMs are typically trained on vast amounts of diverse data from multiple sources~\cite{carlini2021extracting}, some of which might contain inaccuracies, biases, or outdated information. As a result, the model inadvertently produces outputs that are factually incorrect. This phenomenon can be especially problematic in sensitive areas such as healthcare, finance, or legal advice, where incorrect information can have serious consequences.
LLM-generated hallucinations are often perceived as credible and are difficult for users to detect, thereby increasing the likelihood that such misinformation is trusted and propagated, potentially leading to cascading harmful rumors~\cite{nahar2024fakes}.
\begin{examplebox1}
\textbf{Practical Example (LLM\_09): Misinformation Leading to Phishing and Account Takeover.}
An LLM-powered customer-support assistant generates a convincing but false security warning asking users to re-verify their account through a link. An attacker manipulates the assistant using poisoned FAQ content or adversarial prompts so the link points to a phishing page. Users who trust the assistant may enter credentials, enabling account takeover. This shows how LLM-generated misinformation can automate social engineering for classical web attacks.
\end{examplebox1}
\subsubsection{TRA\_09: Content Manipulation and Social Engineering}
In web information systems, attackers often exploit manipulated content to influence user perception and decision-making. Such attacks commonly appear in the form of disinformation campaigns, fake news propagation, phishing messages, or other social engineering techniques where malicious actors distribute content that appears legitimate but is intentionally deceptive. These attacks rely on manipulating the credibility of information rather than exploiting technical software flaws.
\paragraph{Structural Similarity}
In both LLM\_09 and TRA\_09, the attack exploits the human trust layer. In web systems, adversaries present misleading content that users trust and act upon. Similarly, in LLM-based systems, generated outputs are often perceived as legitimate, enabling automated trust exploitation. In both cases, user trust becomes the key vulnerability. However, while traditional web attacks are typically manual and limited in scale, LLM-driven exploitation is automated and can operate at a much larger scale.

\subsubsection{Amplification by LLMs.}
LLMs significantly amplify this risk because they autonomously generate natural language responses that may not always be grounded in verified knowledge sources. Unlike conventional systems where misinformation might originate from external web content, LLMs can internally generate incorrect or fabricated information, often referred to as hallucinations. These outputs may appear credible due to the model’s linguistic fluency and contextual reasoning abilities. Furthermore, adversaries may intentionally manipulate model behavior through prompt injection, adversarial prompting, or data poisoning, causing the model to produce misleading narratives, fabricated facts, and biased explanations. As a result, LLM systems can unintentionally act as large-scale amplifiers of misinformation~\cite{ji2023survey}.

\subsubsection{Current Mitigation Approaches.} 
To mitigate the spread of misinformation in LLM systems, robust fact-checking and validation mechanisms should be integrated into the generation pipeline. These mechanisms include automated systems that verify generated claims against trusted external knowledge  sources~\cite{rahman2026hallucination, chen2024combating}. One promising approach is the ``Attribute First, Then Generate'' framework, which integrates concise and localized source attributions during text generation so that each generated claim is supported by verifiable evidence. Such attribution mechanisms improve transparency and allow end-users to verify the credibility of generated responses by tracing information back to specific source spans~\cite{slobodkin2024attribute}. 
Additionally, LLMs should undergo continuous updates and retraining to ensure access to the most current and reliable information. In high-risk domains, a Human-in-the-Loop (HITL) validation mechanism can also be implemented, where a human evaluator reviews model outputs before they are delivered to users~\cite{amirizaniani2024llmauditor}. 
Future research should therefore focus on developing stronger grounding techniques, reliable knowledge verification mechanisms, and robust evaluation frameworks to better detect and mitigate misinformation generated by LLM systems.
\begin{examplebox2}
\textbf{Open Research Challenges.}
Language models generate responses based on probabilistic reasoning rather than verified factual knowledge, which makes it difficult to guarantee complete factual accuracy. Moreover, balancing factual verification with model efficiency and scalability remains an ongoing research problem, particularly for real-time applications.
\end{examplebox2}

\subsection{\textbf{LLM\_10: Unbounded Consumption}}
\label{sec:llm10}
During inference, LLMs apply learned patterns and knowledge to generate outputs. However, a significant risk arises when systems allow users to issue uncontrolled inference requests, which can lead to various issues such as denial of service (DoS) attacks, economic losses, model theft, and, more commonly, resource exploitation, particularly in cloud-based environments. 
These availability attacks are fundamentally system-level rather than model-level, where adversaries can induce severe latency amplification (up to 280×) at significantly lower cost, thereby highlighting the risk of resource-exhaustion attacks in LLM serving infrastructures~\cite{wang2026rethinking}.
This risk manifests in several ways, including flooding the system with varying lengths of input, continuous input overflows, resource-intensive queries, and model resource extraction via APIs~\cite{fu2024serverlessllm}. Recent overthinking attacks further illustrate this risk, where adversaries craft prompts that induce unnecessarily long reasoning chains, increasing token consumption, latency, and operational cost~\cite{liu2026badthink}. Additionally, model replication and side-channel attacks are other common forms of exploitation that target the system’s resources~\cite{zheng2024inputsnatch}. These attacks can severely compromise the model’s performance, leading to high operational costs and degraded services.
\begin{examplebox1}
\textbf{Practical Example (LLM\_10): Unbounded Consumption Leading to Economic DoS.}
A public-facing travel website uses an LLM chatbot to search bookings, summarize policies, and call backend APIs. An attacker sends long, multi-step prompts across many sessions, forcing repeated inference, retrieval, and API calls. Unlike traditional DoS, each request triggers costly LLM and backend operations, degrading availability, increasing cloud costs, and turning web-layer resource exhaustion into economic denial of service.
\end{examplebox1}
\subsubsection{TRA\_10: Resource Exhaustion and Economic DoS (CWE-400, CWE-834)}
Traditional web systems are vulnerable to attacks that deliberately exhaust computational resources such as CPU, memory, or network bandwidth (CWE-400, where a system fails to properly limit the resources that can be consumed by users or processes), and CWE-834 which describes scenarios where computational processes are forced to execute repeated or overly complex operations, leading to performance degradation or service disruption~\cite{cwe-400,cwe-834}. Such attacks often manifest as DoS attempts that overwhelm servers with resource-intensive requests.
\paragraph{Structural Similarity}
In both LLM\_10 and TRA\_10, the core issue is the exhaustion of system resources, leading to a breakdown in availability. In traditional web systems, attackers overwhelm servers by flooding them with excessive requests, causing service disruption. Similarly, in LLM-based systems, adversaries issue computationally expensive prompts that overload inference resources, resulting in economic or compute-based denial of service. The structural similarity lies in the violation of system availability through resource exhaustion.
\subsubsection{Amplification by LLMs.}
LLM-based systems significantly amplify these risks because generating responses often requires substantial computational resources, including GPU processing, memory allocation, and token generation loops. Attackers often exploit this behavior by crafting extremely long prompts, repeatedly triggering expensive reasoning tasks, or issuing high-frequency API calls. Unlike traditional web systems, DoS attacks that primarily rely on network traffic, LLM abuse involves computational amplification, where a single request can trigger disproportionately large amounts of processing. This makes LLM services particularly vulnerable to resource exhaustion and economic denial-of-service (EDoS) attacks~\cite{wang2026rethinking}.

\subsubsection{Current Mitigation Approaches.}
Resource limiting is an effective defense against unbounded consumption issues. This enforces strict caps on CPU usage, memory allocation, network bandwidth, and token generation limits during inference and training. These controls prevent the model from consuming excessive resources and ensure that operations remain within predefined performance boundaries~\cite{OWASP_LLM10, lakha2025faster}.
Rate limiting also plays a crucial role in regulating the frequency of user requests. By restricting the number of requests processed within a given time window, systems can prevent request flooding and maintain stable resource utilization~\cite{bari2025optimal}.
Model optimization techniques can further reduce computational demands. Approaches such as quantization~\cite{frantar2022gptq} and model pruning~\cite{jia2024model} help reduce model size and inference cost while preserving performance.
In cloud-based environments, dynamic scaling and load balancing help maintain system responsiveness during fluctuating workloads. These mechanisms allocate resources adaptively across distributed systems to prevent individual components from becoming overloaded~\cite{jain2024intelligent}.
Anomaly detection and continuous monitoring are also critical. These techniques help identify abnormal usage patterns such as sudden spikes in CPU utilization, token generation rates, or request frequency, enabling administrators to respond quickly to potential abuse~\cite{lakha2025faster}. 

Finally, cost management and usage monitoring remain essential in large-scale LLM deployments. Cloud-based billing models can make systems vulnerable to economic denial-of-service attacks, where adversaries deliberately trigger expensive computations. Effective cost monitoring mechanisms can detect abnormal spending patterns and prevent unsustainable operational costs~\cite{singh2025optimizing, liagkou2024cost}.
Hence future work should develop cost-aware security frameworks that monitor both resource abuse and computational expenses in real time. Adversarial-prompt anomaly detection, adaptive limits on reasoning depth and tool access, and standardized benchmarks for LLM resource-exhaustion attacks remain important research directions.
\begin{examplebox2}
\textbf{Open Research Challenges.}
Although rate limiting, resource capping, and anomaly detection provide practical defenses, key challenges remain. It is still difficult to distinguish legitimate high-volume use from adversarial resource abuse, detect adversarial prompts that trigger excessive reasoning or tool calls, and control economic denial-of-service risks in cloud-based LLM services. Autonomous agents further amplify this risk by chaining model interactions, API calls, and external tools.
\end{examplebox2}
\subsection{CVE Evidence of Zero-Day-Like LLM-Enabled Web Vulnerabilities}
\label{sec: CVE Evidence of Zero-Day-Like LLM-Enabled Web Vulnerabilities}

Recent entries in the Common Vulnerabilities and Exposures database demonstrate how LLM-mediated vulnerabilities are no longer only
theoretical challenges, but are beginning to appear as concrete, publicly disclosed software and web weaknesses.
For instance, EchoLeak (CVE-2025-32711)~\cite{CVE-2025-32711}  showed how a zero-click vulnerability in Microsoft 365
Copilot enabled sensitive data exfiltration through an LLM scope-violation attack. Similarly, CVE-2024-5826~\cite{CVE-2024-5826} and CVE-2025-1497~\cite{CVE-2025-1497} demonstrate how prompt injection and insufficient validation of LLM-generated code can lead to remote code execution, while CVE-2024-7764~\cite{CVE-2024-7764} illustrates how LLM-generated SQL can re-enable classical SQL injection behavior. CVE-2025-31363~\cite{CVE-2025-31363} further shows how prompt injection in an AI plugin's Jira tool can enable tool-mediated data exfiltration. Recent 2026 cases, such as CVE-2026-27740~\cite{CVE-2026-27740}, show that LLM outputs can trigger stored XSS when downstream web applications trust AI-generated content without proper sanitization. These examples strongly support our central argument that LLM-enabled web systems can transform traditional web vulnerabilities into zero-day-like interaction failures that expand client, server, pipeline, and agent
layers.

\subsection{Answering the Research Questions}

\noindent\textbf{RQ1 (Amplification of traditional vulnerabilities).} Our OWASP–TRA–CWE analysis shows that each OWASP LLM risk maps to a corresponding TRA category, with CWE grounding where applicable, and that LLM integration amplifies these web-security risks rather than replacing them. The recurring mechanism is a shift from syntactic to semantic exploitation: prompt injection re-enables instruction-injection and confused-deputy attacks (TRA\_01) at the level of meaning rather than code, improper output handling turns model text into downstream XSS and SQL injection (TRA\_05), excessive agency converts prompt manipulation into CSRF-style privileged actions (TRA\_06), and embedding weaknesses transform side-channel inference into representation-level leakage (TRA\_08). Across these cases, LLMs act as intermediaries that carry backend and pipeline weaknesses into concrete client-side harms, and the disclosed CVEs in Section~\ref{sec: CVE Evidence of Zero-Day-Like LLM-Enabled Web Vulnerabilities} confirm that these are no longer only theoretical.

\noindent\textbf{RQ2 (Sufficiency of existing defenses and standards).} Existing defenses and standards are necessary but not sufficient. As summarized in Table~\ref{tab:owasp_llm_compact_matrix}, per-risk mitigations exist for each category, yet they leave consistent gaps in data-instruction separation, output-to-action isolation, agent governance, and runtime leakage detection.
These findings motivate the framework-level analysis in Section 6, where we examine RQ2 in detail.

\section{Extending Security Frameworks for LLM-Enabled Web Systems (RQ2)}
Building on the defense gaps identified in Section~5, Table~\ref{tab:framework_limitations} shows that existing frameworks and taxonomies, including OWASP, NIST AI RMF, ISO/IEC standards, and CWE, provide limited coverage of adversarial natural-language manipulation, agentic behavior, and post-deployment behavioral drift. These limitations motivate our LLM-aware monitoring-control framework, which couples governance objectives with concrete control layers across the interaction pipeline.
LLM-enabled systems not only inherit classical vulnerabilities but also fundamentally alter their manifestations by weakening key security boundaries, including data–instruction separation, trust boundaries, and execution control.  Consequently, existing security measures are insufficient to address these interaction-driven risks. 
Several established security frameworks provide structured guidance for managing security risks. These frameworks are being updated to address emerging challenges introduced by AI and LLM systems, such as the National Institute of Standards and Technology AI Risk Management Framework (NIST AI RMF)~\cite{ai2023artificial} and ISO standards discussed below~\cite{international2023iso}.
\subsection{AI Security and Risk Governance using NIST AI RMF and ISO/IEC Standards} 
Security frameworks are important because they translate technical risks, including both web system risks and LLM-mediated risks, into structured governance, mitigation, and monitoring practices. Such frameworks help organizations manage vulnerabilities through access control, secure design, incident response, and continuous risk assessment.
The NIST RMF provides a systematic approach for identifying and mitigating challenges in AI-enabled systems based on four core principles- Govern, Map, Measure, and Manage. These principles guide organizations in identifying potential harms, assessing risks, implementing mitigation strategies, and continuously monitoring the AI systems throughout its life-cycle. 
The framework promotes trustworthy AI by emphasizing reliability, safety, security, transparency, explainability, privacy protection, and fairness~\cite{ai2023artificial, ai2024artificial}.
On the other hand, ISO/IEC AI standards support security and privacy governance through risk assessment, access control, secure data handling, and AI lifecycle management. Therefore, adapting ISO/IEC standards to incorporate LLM-specific auditing, human oversight, and lifecycle security controls is essential for robust governance in AI-integrated environments~\cite{international2023iso, patil2025advancing}.
However, as shown in our analysis, LLM-enabled web systems extend these risks beyond deterministic software flaws to include prompt-based manipulation, agentic actions, prompt leakage, unsafe outputs, and post-deployment behavioral drift.
\subsection{Limitations of Existing Frameworks}
Existing security frameworks, such as OWASP, NIST AI RMF, ISO/IEC standards, and the CWE taxonomy, generally provide a baseline for identifying, categorizing, and understanding cybersecurity risks. Because these frameworks were primarily designed for deterministic and traditional cybersecurity risks, they sometimes fail to fully address the unique challenges and dynamic behavior of LLMs and LLM-enabled applications (see Table~\ref{tab:framework_limitations}). In particular, they lack the ability to model adversarial natural-language manipulation, agent-driven behaviors, and post-deployment risks arising from continuous user–model interaction. To systematically analyze these gaps, we summarize the limitations of existing frameworks and highlight the required extensions for LLM-aware security in Table~\ref{tab:framework_limitations}.

\begin{table*}[t]
\caption{Limitations of Existing Security Frameworks and Weakness Taxonomies in LLM-Enabled Systems.}
\label{tab:framework_limitations}
\centering
\small
\setlength{\tabcolsep}{4pt}
\renewcommand{\arraystretch}{0.95}
\begin{tabularx}{\textwidth}{p{3cm} X X}
\toprule
\textbf{Framework/Database} & 
\textbf{Limitation in LLM Systems} & 
\textbf{Required Extension} \\
\midrule
OWASP & Limited treatment of adversarial natural-language and prompt-level manipulation. & Instruction-aware threat modeling and prompt-level validation. \\
NIST AI RMF & Limited guidance for runtime monitoring and post-deployment behavioral changes. & Continuous AI monitoring, drift detection, and lifecycle adaptation. \\
ISO/IEC & Static policy orientation with limited adaptability to dynamic LLM behavior. & Adaptive, context-aware governance and policy enforcement. \\
CWE & Limited coverage of prompt-level, agentic, and LLM-specific weakness classes. & Extension toward LLM-specific weaknesses such as prompt injection, embedding leakage etc. \\

\bottomrule
\end{tabularx}
\end{table*}
Given the current landscape, where users increasingly interact with LLM-powered systems, security frameworks must evolve to incorporate LLM-aware defenses across multiple layers of the system.  In this direction, Abuadbba et al.~\cite{abuadbba2026human} proposed the 4C framework, which emphasizes the importance of cognitive processes, interaction dynamics, and control mechanisms in securing agentic AI systems, thereby extending beyond traditional web-security and system-centric perspectives.
Furthermore, NIST highlights post-deployment AI monitoring as a major challenge due to behavioral drift, limited standards, and adversarial natural-language interactions~\cite{NIST1}. In this direction, we propose an LLM-aware security framework that combines semantic validation, prompt integrity, output isolation, agent control, behavior, and continuous monitoring for LLM-enabled systems.
\subsection{Proposed LLM-Aware Security Framework Extensions}
Despite their strengths, NIST AI RMF and ISO/IEC frameworks remain limited for LLM-enabled systems because they lack semantic input validation, prompt–execution separation, output-to-action control, agent autonomy governance, and continuous post-deployment monitoring. These gaps make it difficult to address the dynamic, interaction-driven attack surface introduced by LLMs.

\begin{figure}[t]
    \centering
    \includegraphics[width=0.60\textwidth,trim=15 12 15 12,clip]{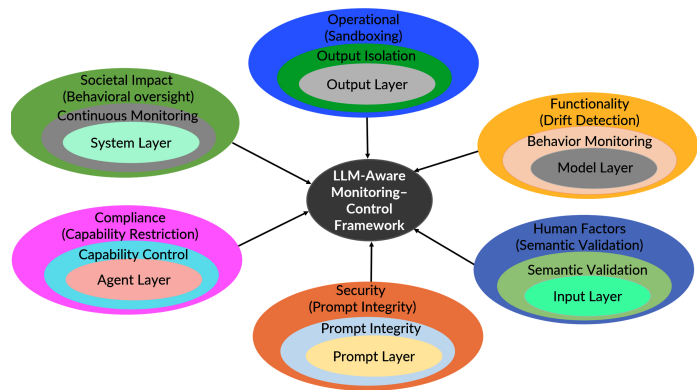} 
   \caption{LLM-aware monitoring–control framework integrating NIST-defined monitoring dimensions (outer nodes) with LLM-specific security controls. Arrows represent continuous post-deployment monitoring and enforcement across system interactions.}
    \label{fig:LLM_Frame}
\end{figure}

Recent work by NIST (AI 800-4)~\cite{NIST1} highlights the importance of post-deployment monitoring and identifies six key monitoring dimensions for AI systems. 
To bridge this gap, we integrate LLM-specific security controls with the NIST monitoring dimensions, providing a unified framework that connects monitoring objectives with actionable defense mechanisms. Specifically, existing frameworks should be extended to incorporate (i) prompt-level integrity verification, (ii) semantic input validation mechanisms, (iii) output-to-action isolation controls, (iv) capability-constrained agent governance, and (v) continuous behavioral monitoring for detecting drift and adversarial manipulation. 
As illustrated in Figure~\ref{fig:LLM_Frame}, each monitoring dimension is coupled with a corresponding LLM-specific control, enabling continuous and behavior-aware security enforcement tailored to the dynamic and interaction-driven nature of LLM systems.
Further,  Figure~\ref{fig:LLM_Frame} operationalizes our proposal by mapping high-level governance objectives to concrete LLM-specific control layers across the interaction pipeline.
For example, \textit{consider an LLM-enabled assistant tool tasked with processing user queries and accordingly interacting with internal APIs (e.g., retrieving reports). According to the proposed framework, received user queries are first processed through a semantic input validation layer (Input Layer), which identifies potentially adversarial instructions. The validated input is then passed to a prompt integrity layer (Prompt Layer), which enforces strict separation between system-level instructions and user-provided content.
At the model level, behavioral monitoring (Model Layer) continuously analyzes response patterns to detect anomalies. Before any model-generated output is executed, it passes through an output-to-action isolation layer (Output Layer), ensuring that responses are treated as untrusted data and cannot directly trigger system operations without verification.
Simultaneously, agent-level capability control (Agent Layer) restricts the assistant’s access to sensitive resources (e.g., financial or medical data), enforcing least-privilege execution. Finally, continuous system-level monitoring (System Layer) tracks interaction patterns over time to detect drift, misuse, or coordinated attacks.}

This layered integration ensures that each stage of the LLM interaction pipeline is governed by explicit security controls. As shown in Figure~\ref{fig:LLM_Frame}, governance objectives are mapped to concrete control layers, translating high-level risk principles into actionable defenses for LLM-enabled systems.




\subsection{\textbf{Threats to Validity}}
\noindent\textbf{Construct validity.} Our analysis rests on the mapping between OWASP LLM risks, the TRA categories defined in this work, and CWE entries. Because this mapping involves interpretation, different researchers could assign a given risk to a different traditional analogue. To reduce this threat, we derived each mapping from four explicit criteria (entry point, processing mechanism, security boundary violation, and impact) described in Section 3.3, and grounded every TRA category in established CWE entries where applicable, rather than in surface-level analogy.

\noindent\textbf{Internal validity.} The study selection may have missed relevant work, since our search keywords were seeded from the OWASP LLM Top 10 and traditional attack names. We mitigated this through forward and backward snowballing and reference analysis, which added studies beyond the initial keyword search. The coding of the 105 primary studies (addressed risk, affected layer, and proposed mitigation) was performed by a single author. Consistency was maintained by coding every study against the fixed OWASP-TRA-CWE scheme and by re-reviewing all assignments in a second pass, but we acknowledge single-coder bias as a limitation. In addition, a portion of the primary studies are arXiv preprints that have not undergone peer review. Given the pace of LLM security research, excluding preprints would omit influential recent results; we therefore included them selectively based on citation impact and relevance, while noting that their findings may change upon formal publication.

\noindent\textbf{External validity.} Our conclusions are tied to the 2025 OWASP LLM Top 10 as the organizing lens, and both the risk list and the threat landscape evolve quickly. The amplification patterns and framework gaps we identify reflect the literature and the disclosed CVEs available at the time of the search, and new attack classes may emerge that fall outside the ten categories analyzed here. We believe the TRA-based analysis method itself generalizes, since it links risks to root-cause weaknesses rather than to a specific list version, and it can be reapplied as OWASP and CWE are updated.
\section{Conclusion}
In this paper, we examined how web security risks, such as XSS and CSRF, are re-enabled and amplified in LLM-driven systems. We introduced a taxonomy that maps these risks to the OWASP LLM Top 10 and analyzed how LLM-related threats emerge across client-side, server-side, and pipeline layers. Our analysis shows that LLM-mediated interactions weaken data–instruction, trust, and execution boundaries, creating a dynamic and interaction-driven attack surface. Based on these findings and the limitations of existing mitigation approaches, we proposed an LLM-aware monitoring–control framework that connects governance objectives with concrete application- and system-level defenses. Overall, this survey shows that LLMs fundamentally transform the web threat landscape and highlights the need for robust defenses to protect users from both legacy and emerging security threats.

\bibliographystyle{ACM-Reference-Format}
\bibliography{References}
\clearpage





\end{document}